\documentclass[11pt,a4paper]{article}
\usepackage{jheppub}
\usepackage[T1]{fontenc}
\usepackage{lmodern}
\usepackage{tikz}
\usetikzlibrary{arrows.meta,decorations.markings}
\newcommand{\dd}{\mathrm{d}}
\newcommand{\ii}{\mathrm{i}}
\renewcommand{\Re}{\operatorname{Re}}
\definecolor{contourblue}{HTML}{0077BB}
\definecolor{cutorange}{HTML}{CC3311}
\allowdisplaybreaks[1]
\compress
\preprint{{\footnotesize USTC-ICTS/PCFT-26-71}}
\title{IR extremality and UV softness in string bootstrap}
\author[a]{Shi-Lin Wan,}
\author[b]{Ji-Yang Wang,}
\author[a,c]{and Shuang-Yong Zhou}
\affiliation[a]{Interdisciplinary Center for Theoretical Study, University of Science and Technology of China,\\Hefei, Anhui 230026, China}
\affiliation[b]{Department of Modern Physics, University of Science and Technology of China,\\Hefei, Anhui 230026, China}
\affiliation[c]{Peng Huanwu Center for Fundamental Theory,\\Hefei, Anhui 230026, China}
\emailAdd{wsl9868@mail.ustc.edu.cn}
\emailAdd{wangjiyang@mail.ustc.edu.cn}
\emailAdd{zhoushy@ustc.edu.cn}
\abstract{
Whether string amplitudes can be characterized by extremality within the space of consistent S-matrices is an intriguing question.
We address this question within the framework of positivity bootstrap. We compute positivity bounds by minimizing the $F^4$ Wilson coefficient while fixing subleading coefficients, and show that the Veneziano amplitude lies close to the boundary of the allowed region. However, this numerical proximity does not imply exact extremality, as demonstrated by explicit counterexamples. When additional KLT-motivated UV softness is imposed, the Veneziano amplitude is found to uniquely minimize the $F^4$ coefficient for $4\leq D\leq10$, connecting the extremality of the Veneziano amplitude to its stringy UV behavior and offering a new perspective on what makes string theory special.
}

\begin{document}
\maketitle

\section{Introduction}

Determining which low-energy effective theories admit consistent ultraviolet (UV) completions is a central problem in high-energy physics.
In quantum gravity, this question motivates the swampland program \cite{Vafa:2005ui,Palti:2019pca}.
The S-matrix bootstrap approaches UV consistency directly through crossing symmetry, analyticity, locality, and unitarity of scattering amplitudes, without specifying a microscopic Lagrangian \cite{deRham:2022hpx,Kruczenski:2022lot}.
Representative analytic and numerical developments include nonperturbative bounds on scattering amplitudes \cite{Paulos:2016but,Paulos:2017fhb,Karateev:2019ymz,Guerrieri:2021tak,EliasMiro:2022xaa,deRham:2025vaq}, systematic constraints on EFT Wilson coefficients \cite{Adams:2006sv,deRham:2017avq,Remmen:2020vts,Zhang:2020jyn,Bellazzini:2020cot,Tolley:2020gtv,Caron-Huot:2020cmc,Arkani-Hamed:2020blm,Sinha:2020win,Li:2021lpe,Chiang:2021ziz,Chiang:2022ltp,Cheung:2025nhw,Chang:2026ztn}, and methods for UV reconstruction \cite{Wan:2024eto,Calisto:2026pvv}.
Applications of these methods and their variants include gauge theories and gravity \cite{Bern:2021ppb,Caron-Huot:2021rmr,Caron-Huot:2022ugt,Henriksson:2022oeu,Fernandez:2022kzi,Ma:2023vgc,He:2023lyy}, as well as string scattering \cite{Huang:2020nqy,Chiang:2023quf,Haring:2023zwu,Berman:2023jys,Berman:2025owb}.

In string bootstrap, analytic studies have connected string-like asymptotic spectra to high-energy behavior \cite{Caron-Huot:2016icg}.
Exact uniqueness results supplement consistency with level-truncation or ultrasoft Regge behavior and minimal residue zeros for the Veneziano amplitude \cite{Cheung:2024uhn, Cheung:2025tbr}, or the string monodromy or splitting condition for the Veneziano amplitude \cite{Wan:2026pjq,Chiang:2023quf,Berman:2023jys,Berman:2025owb}.
A complementary route combines maximal supersymmetry with higher-point factorization, R-symmetry, scalar parity, and positivity \cite{Elvang:2026pmc,Huang:2026ghg,Chiang:2026ubv,Berman:2026ezk}.
Related EFT studies also use higher-point consistency to derive string monodromy relations and constrain the KLT kernel \cite{Chen:2022shl,Chen:2023dcx}, while parallel investigations address the bootstrap of closed-string amplitudes \cite{Geiser:2022exp,Cheung:2024obl,Cheung:2025tbr,Berman:2026ezk}.

Extremizing low-energy observables within the bootstrap provides another way to distinguish string amplitudes from the broader class of consistent theories.
Nonperturbative S-matrix bootstrap of maximally supersymmetric graviton scattering finds that the smallest allowed $R^4$ coefficient is close to the minimum values realized in string theory and M-theory \cite{Guerrieri:2021ivu,Guerrieri:2022sod}.
At tree level, positivity-bootstrap studies in maximally and half-maximally supersymmetric theories provide evidence for extremal amplitudes supported on linear Regge trajectories, while the Type II string amplitude lies in the interior of the allowed region \cite{Albert:2024yap,Boisvert:2026sfh}.
For open-superstring scattering, imposing a spin-0 state at the first massive pole with a string-inspired mass gap to heavier states produces a sharp corner near the Veneziano point, while additionally fixing its coupling relative to the $F^4$ coefficient to the string value isolates a small allowed island around it \cite{Berman:2024wyt}. Ref~\cite{Bhat:2024agd} found solutions close to the open-superstring amplitude by minimizing the $F^4$ Wilson coefficient $A_{0,0}$, with an equally spaced mass-squared spectrum and two higher-order Wilson coefficients fixed to their string values. It motivates the minimization objective through the first finite moment of entangling power, with this objective proportional to $A_{0,0}$.

In this paper, we first use fixed-$u$ dispersion relations and the moment-problem method to compute positivity bounds in $D=10$, assuming an equally spaced mass-squared spectrum, minimizing $A_{0,0}$ and fixing additional Wilson coefficients. The minima obtained at finite truncation are remarkably close to the Veneziano value, in agreement with Ref~\cite{Bhat:2024agd}.
Further numerical tests indicate that the Veneziano amplitude remains close to the boundary of the allowed region even when the assumption of an equally spaced mass-squared spectrum is relaxed.
As the truncation order increases, the numerical bounds appear to approach the Veneziano value, albeit slowly.
These observations suggest that the Veneziano amplitude might be the exact minimizer with the prescribed Wilson data held fixed.

However, we construct 2 explicit counterexamples showing that the Veneziano amplitude does not minimize $A_{0,0}$ with the prescribed Wilson data held fixed.
Both preserve the Veneziano mass gap and equally spaced mass-squared spectrum while lowering $A_{0,0}$.
The first keeps the $W_{0,1}$ coefficient at its Veneziano value, while the second also fixes the $A_{1,0}$ coefficient to its Veneziano value.
For both examples, we prove partial-wave positivity at every mass level and spin for $4\leq D\leq10$.
Thus the specified Wilson data do not suffice for exact Veneziano extremality, even within the restricted spectral class.

Nevertheless, the striking numerical proximity to Veneziano suggests that minimizing $A_{0,0}$ captures important features of string scattering and motivates the search for additional input to uniquely identify the Veneziano amplitude.
Guided by the Kawai--Lewellen--Tye (KLT) relation \cite{Kawai:1985xq}, we impose forward ultrasoftness, requiring exponential suppression of $\mathcal A(s,u)$ at large $|s|$ away from the real $s$ axis, for fixed $t=-s-u$ near $t=0$:
\begin{align}
e^{\ii\pi\alpha's}\mathcal A(s,-s-t)\to0,\qquad |s|\to\infty.
\end{align}
The counterexamples feature power-law decays and therefore lie outside this UV class. The exponential decay rate also gives an intrinsic UV scale $\alpha'$, which fixes the string mass gap.
In the positive dispersive bootstrap framework with the massless contribution normalized to $-1/(su)$, this condition is found to imply the following analytical bound
\begin{align}
A_{0,0}\geq(\alpha')^2\zeta(2).
\end{align}
Saturation of this inequality requires the spectral measures to be supported on the zero set of its kernel, selecting the equally spaced mass-squared spectrum and the corresponding spin parity. These spectral constraints yield the string monodromy condition $\mathcal A(s,t)+e^{\ii\pi\alpha's}\mathcal A(s,u)+e^{-\ii\pi\alpha't}\mathcal A(t,u)=0$. It then follows from Ref~\cite{Wan:2026pjq} that equality holds if and only if the amplitude is Veneziano for $4\leq D\leq10$.

Together, these results deepen our understanding of the extremal role of string amplitudes within the space of consistent S-matrices. They also clarify the role of UV behavior in Veneziano extremality and show how saturation of a low-energy bound determines its characteristic spectrum, spin parity, and monodromy.

The paper is organized as follows. Section~\ref{sec:setup} presents the common bootstrap setup and the numerical problem motivating this study.
Section~\ref{sec:counterexamples} constructs the finite-data counterexamples and proves their positivity.
Section~\ref{sec:analytic} develops the conditional extremal theorem, beginning with the KLT motivation before stating forward ultrasoftness, proving the bound and uniqueness, and relating the bound to the zero structure of the forward amplitude.
We summarize the results in Section~\ref{sec:summary}.
Appendix~\ref{sec:appendix-positivity} contains the technical details of the all-level partial-wave positivity proof for our counterexamples. Appendix~\ref{sec:forward-reconstruction} proves the zero-and-pole structure used in Section~\ref{sec:minimal-zeros}.

\section{Bootstrap setup and numerical analysis}
\label{sec:setup}

To make the question of Veneziano extremality precise, we formulate the minimization of the leading $F^4$ Wilson coefficient within the S-matrix bootstrap.
We first present the dispersion relation and unitarity constraints, and then numerically minimize this coefficient while fixing the mass gap and selected higher-order Wilson coefficients.

\subsection{Bootstrap framework}

Let us review the positivity bootstrap that we use, in which crossing symmetry, analyticity, and the positivity part of unitarity are used to constrain scattering amplitudes without specifying an underlying microscopic theory \cite{deRham:2022hpx}.
Specifically, dispersion relations translate the positivity of spectral weights, together with crossing symmetry, into constraints on low-energy Wilson coefficients \cite{Tolley:2020gtv, Caron-Huot:2020cmc, Arkani-Hamed:2020blm, Chiang:2021ziz}.

Consider a color-ordered four-point amplitude $\mathcal A(s,u)$ for massless gauge bosons, stripped of its color and kinematic prefactors.
We use the mostly-plus metric and take all external momenta to be incoming, so that $p_1+p_2+p_3+p_4=0$.
The Mandelstam invariants are
\begin{align}
s=-(p_1+p_2)^2,\qquad t=-(p_1+p_3)^2,\qquad u=-(p_1+p_4)^2,\qquad s+t+u=0.
\end{align}
The all-incoming color ordering $(1,2,3,4)$ has adjacent-channel invariants $s$ and $u$.
The forward scattering limit is when $t=0$, equivalently $u=-s$.
As part of the bootstrap assumptions, we impose real analyticity, $\mathcal A(s^*,u^*)=\mathcal A(s,u)^*$, and planar $s\leftrightarrow u$ crossing symmetry, $\mathcal A(s,u)=\mathcal A(u,s)$.
Throughout this paper, we normalize the massless pole contribution from tree-level Yang-Mills theory to $-1/(su)$.
We also assume the standard fixed-$u$ analytic structure, with poles and cuts associated with physical intermediate states.
Stripping the kinematic prefactor removes two powers of $s$ from the corresponding component amplitude at fixed $u$, so the usual Froissart--Martin-type growth condition on that amplitude translates into\footnote{The Froissart--Martin bound is rigorously established for massive scattering with a nonzero crossed-channel mass gap \cite{Froissart:1961ux,Martin:1965jj}.
For massless amplitudes, an analogous growth condition is commonly adopted in dispersive analyses.}
\begin{align}
\mathcal A(s,u) / s^0 \longrightarrow 0,\qquad |s|\longrightarrow\infty,\qquad u<0\ \text{fixed},
\end{align}
along large contours avoiding the physical singularities.
The contour contribution at infinity thus vanishes, and Cauchy's theorem together with the partial-wave expansion of the discontinuity gives the unsubtracted dispersion relation
\begin{align}
\mathcal A(s,u)=-\frac{1}{s u}+\sum_{\ell=0}^{\infty}\int_{\Lambda^2}^{\infty}\frac{\overline C_\ell^{(\nu)}\!\left(1+2u/s'\right)}{s'-s}\rho_\ell(s')\,\dd s',
\end{align}
where $\nu=(D-3)/2>0$, $D$ is the spacetime dimension, and $\Lambda^2>0$ is a lower bound on the massive spectral support.
The function $\overline C_\ell^{(\nu)}(z)\equiv C_\ell^{(\nu)}(z)/C_\ell^{(\nu)}(1)$ is the normalized Gegenbauer polynomial of degree $\ell$ and parameter $\nu$.
We absorb the positive factor $C_\ell^{(\nu)}(1)$ into $\rho_\ell(s')$ and impose the positivity part of partial-wave unitarity through $\rho_\ell(s')\geq0$.

For convenience, we write $\widehat{\mathcal A}(s,u)\equiv\mathcal A(s,u)+1/(su)$ and express its low-energy expansion in two equivalent forms:
\begin{align}
\widehat{\mathcal A}(s,u)&=\sum_{m=0}^{\infty}\sum_{n=0}^m A_{m,n}s^{m-n} u^n=-\sum_{p,q\geq0}W_{p,q}(s+u)^p(su)^q.
\label{eq:eft}
\end{align}
Crossing implies $A_{m,n}=A_{m,m-n}$ and permits the expansion in the symmetric variables $s+u$ and $su$.
The two coefficient conventions are related at the first few orders by
\begin{align}
W_{0,0}=-A_{0,0},\qquad W_{1,0}=-A_{1,0},\qquad W_{0,1}=2A_{2,0}-A_{2,1}.
\end{align}
It is usually assumed in dispersive positivity analyses that the low-energy expansion can be interchanged with the spin sum and spectral integral.\footnote{This interchange is justified in the standard gapped setting with suitable analyticity and convergence of the relevant dispersive moments.
A notable obstruction arises in gravity, where the massless $t$-channel pole makes the forward limit singular and requires a separate treatment \cite{Tokuda:2020mlf, Alberte:2021dnj, deRham:2022gfe, Caron-Huot:2021rmr,Caron-Huot:2022ugt, Haring:2024wyz}.}
We thus obtain the dispersive sum rules
\begin{align}
A_{m,n}=\frac{2^n}{n!}\sum_{\ell=n}^{\infty}\int_{\Lambda^2}^{\infty}\frac{1}{(s')^{m+1}}\left.\frac{\dd^n}{\dd z^n}\overline C_\ell^{(\nu)}(z)\right|_{z=1}\rho_\ell(s')\,\dd s'~,~~~0\leq n\leq m~.
\label{eq:amn}
\end{align}
Throughout this paper, our optimization objective is the constant term
\begin{align}
A_{0,0}=\sum_{\ell=0}^{\infty}\int_{\Lambda^2}^{\infty}\frac{\rho_\ell(s')}{s'}\,\dd s'<\infty.
\label{eq:a00}
\end{align}
For the four-gluon component, restoring the stripped kinematic prefactor identifies $A_{0,0}$ with the Wilson coefficient of the leading local $F^4$ interaction in the corresponding color ordering.

In our conventions, the open-superstring Veneziano amplitude with Regge slope $\alpha'$ takes the form
\begin{align}
\mathcal A_{\rm V}^{\alpha'}(s,u)=-(\alpha')^2\frac{\Gamma(-\alpha's)\Gamma(-\alpha'u)}{\Gamma(1-\alpha'(s+u))}.
\end{align}
When the superscript is unnecessary and omitted, we use $\mathcal A_{\rm V}$ to denote $\mathcal A_{\rm V}^{\alpha'}$.
Its massless contribution is $-1/(su)$, and its leading regular coefficient is $A_{0,0}^{\rm V}=(\alpha')^2\zeta(2)$.
In the forward limit $t=0$, it reduces to
\begin{align}
\mathcal A_{\rm V}^{\alpha'}(s,-s)=\frac{\pi\alpha'}{s\sin(\pi\alpha's)}.
\end{align}

\subsection{Numerical methods and results}
\label{sec:numerical}

Numerical positivity bootstrap calculations based on dispersive sum rules bound low-energy coefficients by optimizing over positive spectral measures subject to crossing symmetry.
The sum rules express Wilson coefficients as linear functionals of the spectral measures, while crossing supplies extra linear relations, often organized as null constraints, which are needed to get two-sided bounds \cite{Tolley:2020gtv,Caron-Huot:2020cmc}.
Fixing selected coefficients adds further linear equalities, and minimizing a target coefficient becomes a convex optimization problem.
Discretizing the mass spectrum and retaining finitely many spins and constraints turns the problem into a linear program.
Alternatively, one can search for dual functionals whose kernels are non-negative on the allowed spectral support and thereby bound the objective in the opposite direction.

The same optimization problem can also be formulated as a moment problem \cite{Arkani-Hamed:2020blm,Chiang:2021ziz,Wan:2024eto}.
In this formulation, the optimization variables are moments of the positive spectral measures, and the Wilson coefficients are linear combinations of these moments.
Positivity and restrictions on the spectral support give positive-semidefinite conditions on Hankel matrices and weighted Hankel matrices.
The matrix entries are linear in the moments.
Crossing symmetry and the prescribed Wilson coefficients impose linear constraints on these moments.
At a finite truncation order, the optimization problem becomes a semidefinite program (SDP), and only a subset of the full moment constraints is imposed.
Increasing the truncation order introduces higher moments and larger Hankel matrices, which can further restrict the allowed low-energy coefficients and tighten the bounds.
For a fixed finite set of null constraints, flat-extension theorems provide a criterion for when no further increase in the moment order is needed \cite{Wan:2024eto}.
Once the optimal moment sequence admits a flat extension consistent with the spectral support, it has a positive representing measure, certifying the optimal bound without the need to solve larger SDPs.

Veneziano extremality has been numerically examined previously using a local crossing-symmetric dispersion relation \cite{Bhat:2024agd}, building on representations with an auxiliary parameter $\lambda$ developed in Ref~\cite{Bhat:2025zex}.
Crossing symmetry is manifest in that approach, while $\lambda$-independence supplies linear constraints on the positive partial-wave coefficients after truncation of the mass spectrum.
The parametric relation recovers the usual fixed-variable dispersion relations for suitable choices of $\lambda$, and parameter independence encodes the corresponding null constraints \cite{Saha:2024qpt}.
Here we revisit the numerical analysis using the fixed-$u$ dispersion relation and adopting the moment-problem formulation described above.

Let us now restate the setup, optimization problem, and numerical results of Ref~\cite{Bhat:2024agd} in our notation.
Their open-string bootstrap in $D=10$ uses an equally spaced mass-squared spectrum, with massive poles at $s=N$, $N=1,2,\ldots$, and spins $0\leq\ell\leq N-1$ at level $N$.
The gauge coupling is held fixed, corresponding to our normalization of the massless contribution to $-1/(su)$.
In our conventions, the two prescribed Wilson coefficients are
\begin{align}
A_{1,0}=\zeta(3),\qquad W_{0,1}\equiv2A_{2,0}-A_{2,1}=\frac{7}{4}\zeta(4).
\end{align}
Their minimization objective is the first finite moment of entangling power.
In the unsubtracted dispersive formulation used here, this spectral moment is proportional to $A_{0,0}$, so we shall study the minimization directly in terms of this Wilson coefficient.
After truncating the mass spectrum, they obtain amplitudes close to the Veneziano amplitude, including good agreement along the leading and subleading Regge trajectories in the analysis with the open-superstring spin parity.
This motivates the question: does minimizing $A_{0,0}$ with these inputs select the Veneziano amplitude exactly?

\begin{figure}[h]
  \centering
  \begin{minipage}{0.6\textwidth}
    \centering
    \includegraphics[width=\linewidth,height=0.6\textheight,keepaspectratio]{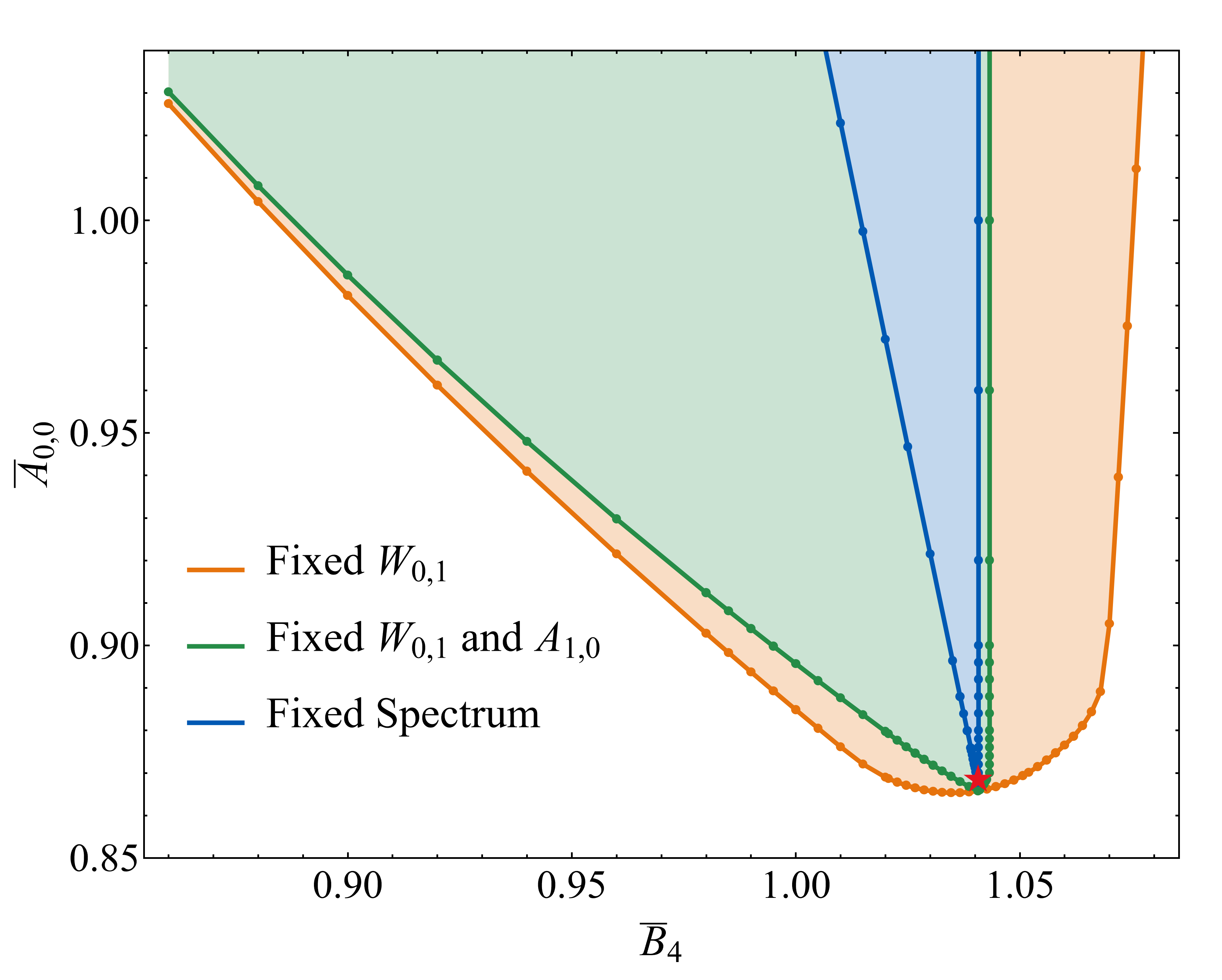}
  \end{minipage}
  \caption{Numerical bootstrap bounds at truncation order $K=21$ in the
  $(\bar{B}_4,\bar{A}_{0,0})$ plane, where
  $\bar{B}_4=B_4/W_{0,1}$ and
  $\bar{A}_{0,0}=A_{0,0}/W_{0,1}$, for three choices of input:
  fixing $W_{0,1}=7\zeta(4)/4$ (orange), fixing both $W_{0,1}$ and
  $A_{1,0}=\zeta(3)$ (green), and imposing the finite low-lying spectrum specified in the main text with both coefficients fixed (blue).
  Shading indicates the regions allowed by the retained constraints. The red star marks the Veneziano amplitude.\\[-21pt]}
  \label{fig:comparison}
\end{figure}

We compare the moment-approach bounds on $A_{0,0}$ with the corresponding Veneziano value $\zeta(2)$.
Exploratory scans suggested a possible kink, motivating us to focus on the coefficient $B_4$, which is defined by
\begin{align*}
\widehat{\mathcal A}(s,-s)=A_{0,0}+W_{0,1}s^2+B_4s^4+O(s^6).
\end{align*}
We study the 2D bounds on $A_{0,0}$ and $B_4$ throughout the rest of the subsection.
The plotted coordinates are $\bar{B}_4=B_4/W_{0,1}$ and $\bar{A}_{0,0}=A_{0,0}/W_{0,1}$; at each chosen $\bar{B}_4$, we optimize $\bar{A}_{0,0}$ subject to the retained Wilson coefficients and crossing constraints.
The Veneziano reference point is
$\left(\bar{B}_4^{\rm V},\bar{A}_{0,0}^{\rm V}\right)
=\Big(31\pi^2/(294),60/(7\pi^2)\Big), ~B_4^{\rm V}=31\zeta(6)/16$.
In the figure, $K$ denotes the truncation order, $m\leq K$, of the null constraints $A_{m,n}-A_{m,m-n}=0$ imposed by crossing symmetry.

\begin{figure}[h]
  \centering
  \begin{minipage}{0.95\textwidth}
    \centering
    \begin{minipage}[t]{0.49\linewidth}
      \centering
      \includegraphics[width=\linewidth,height=0.28\textheight,keepaspectratio]{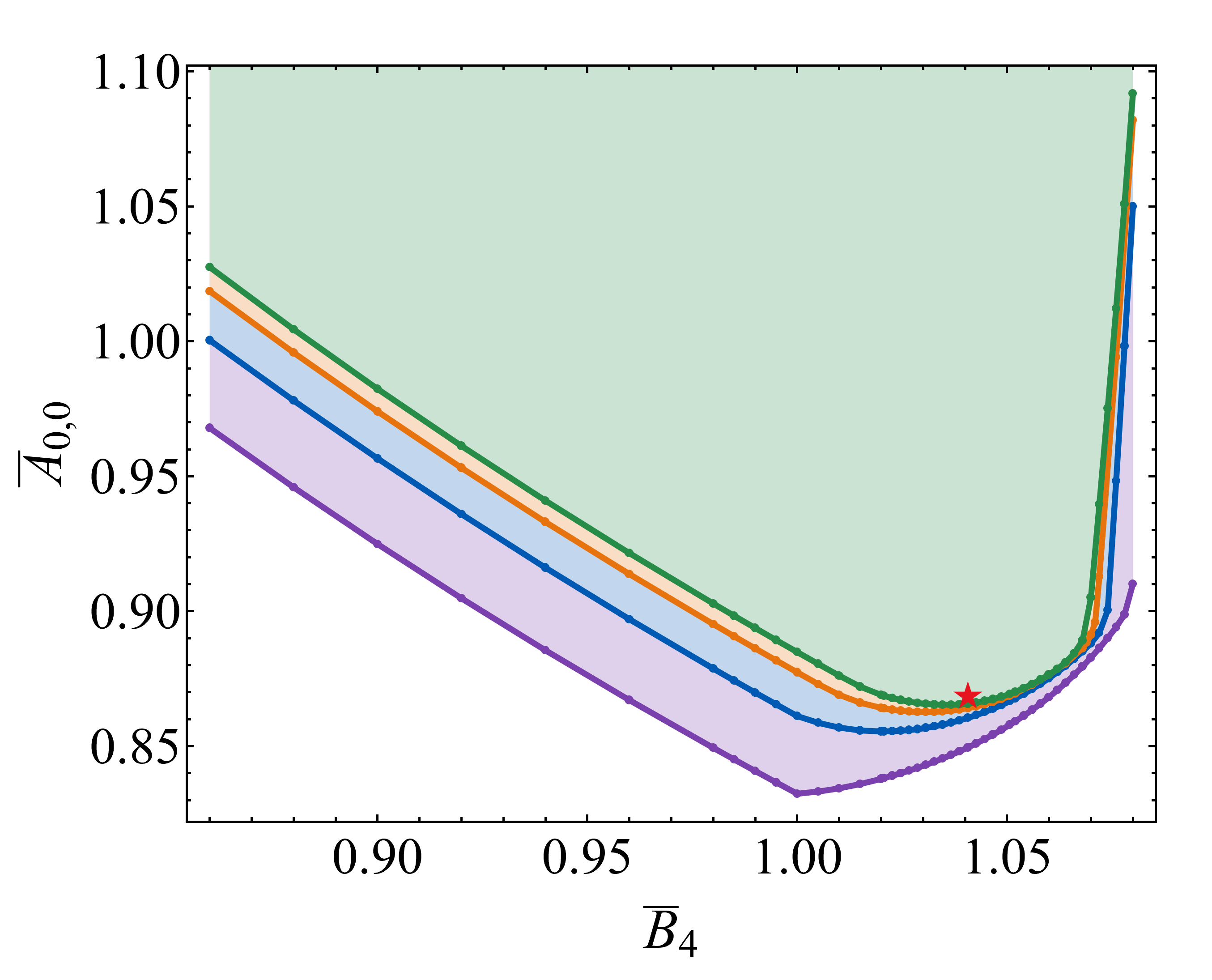}\par
      \smallskip\textnormal{(a)}
    \end{minipage}\hfill
    \begin{minipage}[t]{0.49\linewidth}
      \centering
      \includegraphics[width=\linewidth,height=0.28\textheight,keepaspectratio]{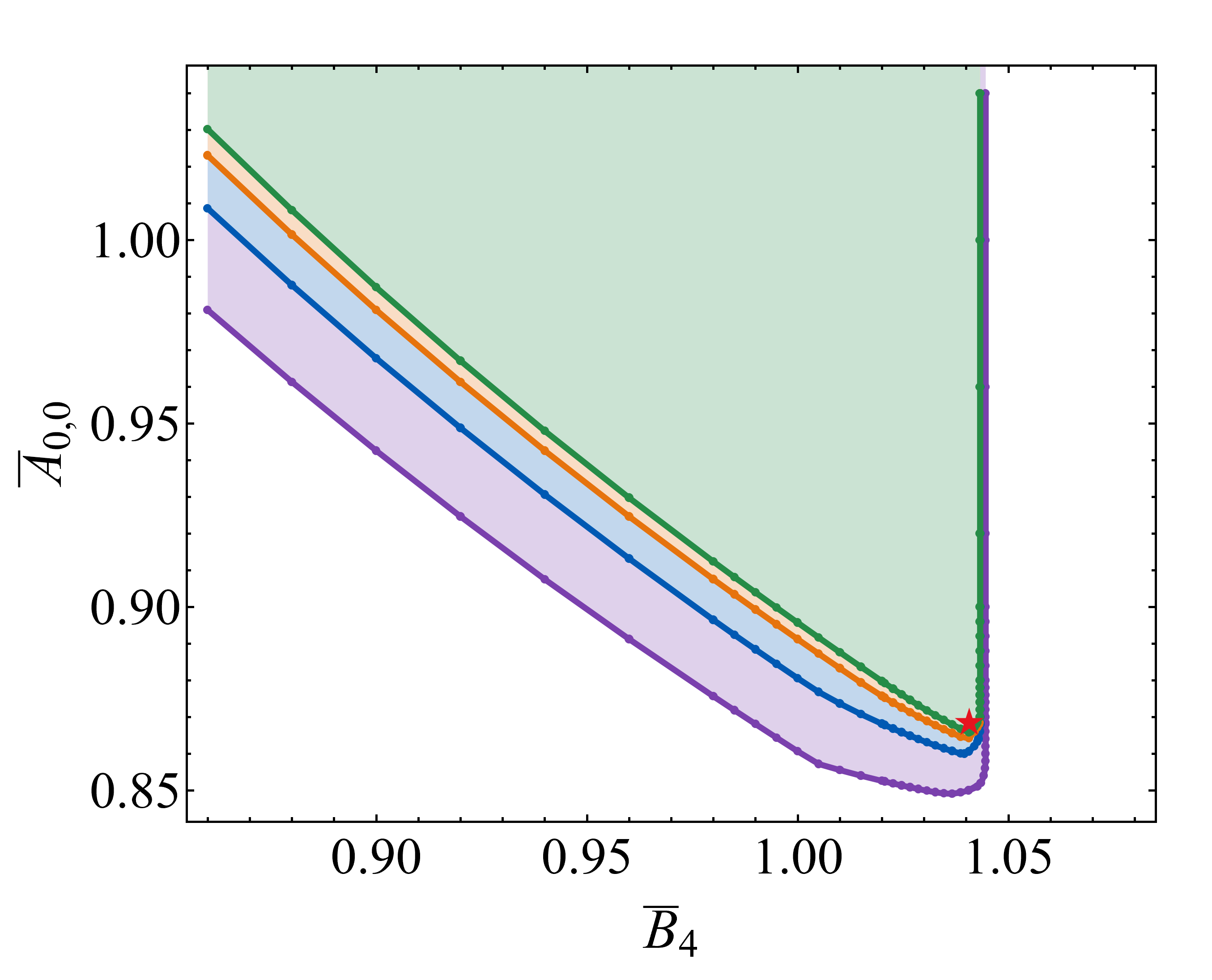}\par
      \smallskip\textnormal{(b)}
    \end{minipage}
    \par\vspace{8pt}
    \begin{minipage}[t]{0.49\linewidth}
      \centering
      \includegraphics[width=\linewidth,height=0.28\textheight,keepaspectratio]{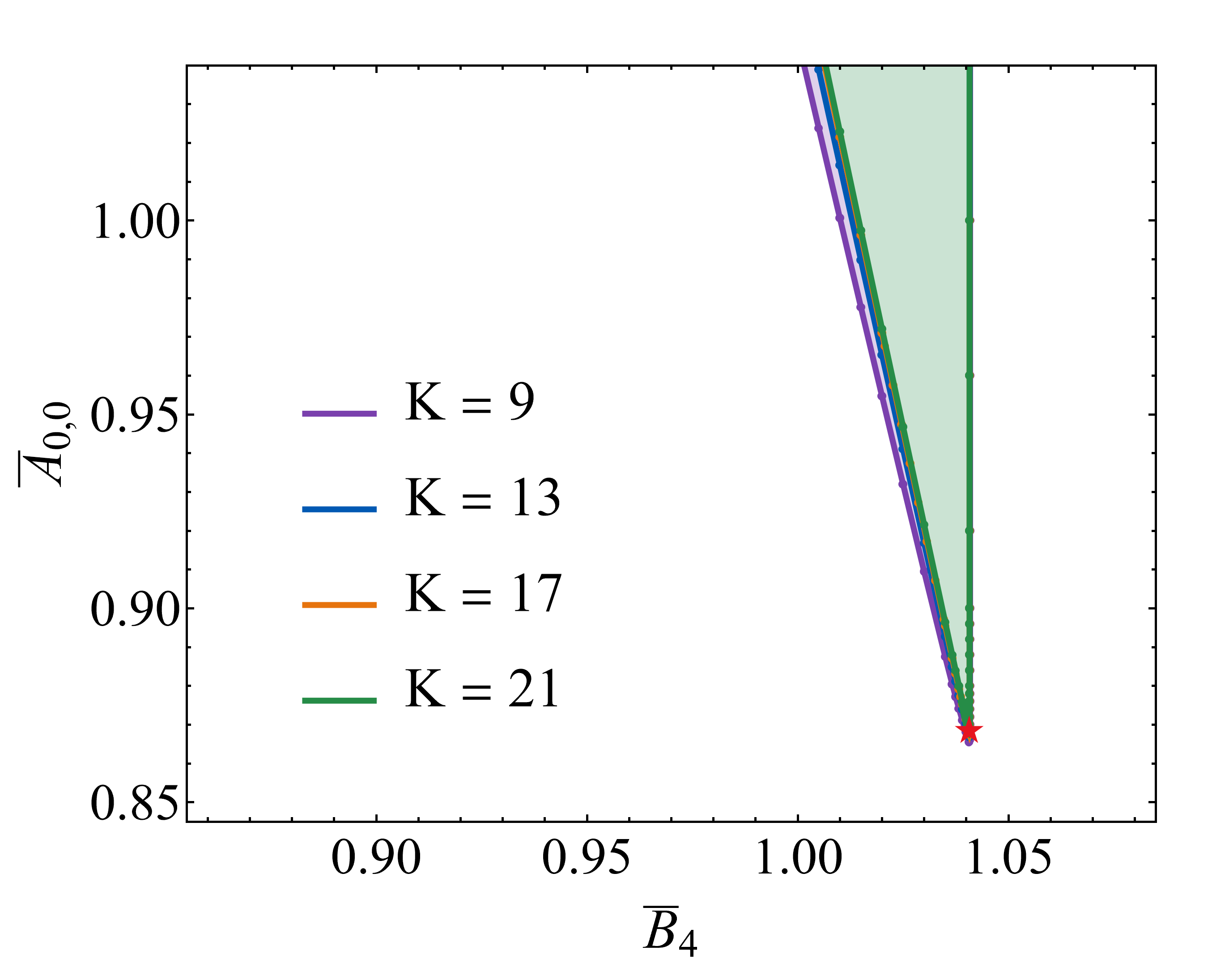}\par
      \smallskip\textnormal{(c)}
    \end{minipage}\hfill
    \begin{minipage}[t]{0.5\linewidth}
      \centering
      \includegraphics[width=\linewidth,height=0.28\textheight,keepaspectratio]{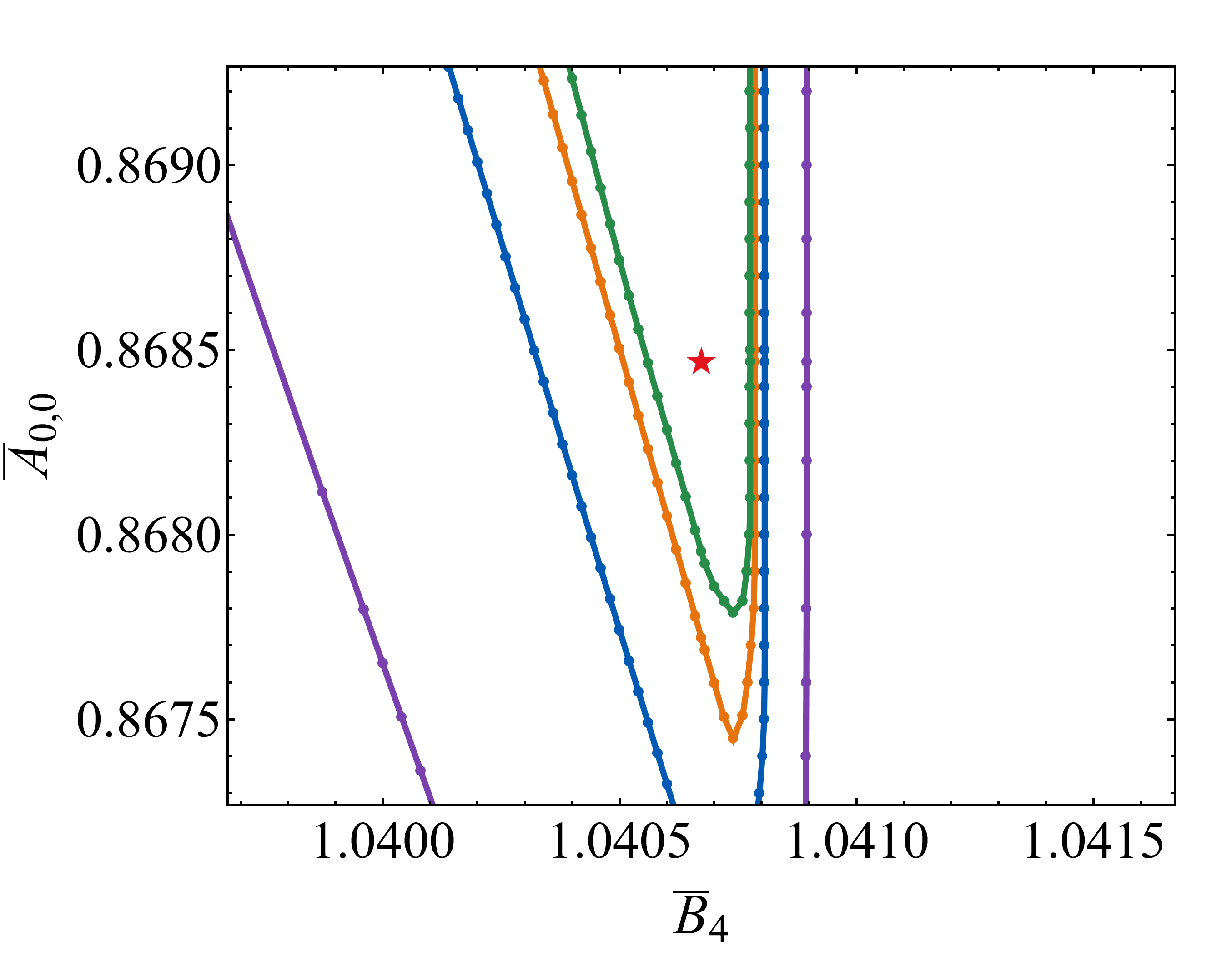}\par
      \smallskip\textnormal{(d)}
    \end{minipage}
  \end{minipage}
  \caption{Truncation dependence of the numerical bootstrap bounds in the
  $(\bar{B}_4,\bar{A}_{0,0})$ plane.
  All panels fix $W_{0,1}=7\zeta(4)/4$.
  (a) $A_{1,0}$ is left unconstrained.
  (b) $A_{1,0}=\zeta(3)$ is also fixed.
  (c) Both Wilson coefficients are fixed, with the finite low-lying spectral input specified in the main text.
  The curves correspond to truncation orders $K=9,13,17,21$;
  shading indicates the regions allowed by the retained constraints.
  (d) An enlarged view of panel (c) near the Veneziano point, using the same colors as in panel (c).
  The red stars mark the Veneziano amplitude.\\[-22pt]}
  \label{fig:truncation-dependence}
\end{figure}

We use a mixed-variate moment method as in Ref~\cite{Wan:2024eto}: we split the spin sum in Eq.~\eqref{eq:amn} at $\ell=\ell_M$;
each partial wave with $\ell<\ell_M$ is treated separately through univariate moments in $x=1/s'$, while the combined contribution from all the higher spins, $\ell\geq\ell_M$, is described by a bivariate moment relaxation in $(x,y)=(1/s', \ell(\ell+D-3)/s')$.
For numerical efficiency, we choose $\ell_M=8$ for $K\leq15$ and $\ell_M=(K+1)/2$ otherwise.

We compare three different inputs: (a) fix $W_{0,1}=7\zeta(4)/4$;
(b) fix $W_{0,1}=7\zeta(4)/4$ and $A_{1,0}=\zeta(3)$;
(c) fix $W_{0,1}=7\zeta(4)/4, ~A_{1,0}=\zeta(3)$ and impose finite low-lying spectral input.
For the third case, the low-lying spectrum is restricted to levels $s'=1,\ldots,19$, with $0\leq\ell\leq N-1$ and $\ell\equiv N-1\pmod2$, while $s'\geq\ell+1$ for the high-spin bivariate part.
The label ``Fixed Spectrum'' in Figure~\ref{fig:comparison} refers to this third choice of constraint.

Figure~\ref{fig:comparison} shows that the numerical proximity to Veneziano persists even when the spectral or Wilson coefficient inputs are relaxed.
This persistence under weaker assumptions strengthens the motivation for seeking an analytic explanation of the Veneziano minimization.
Hints of a kink near Veneziano also motivated our $B_4$ scans.
Figure~\ref{fig:truncation-dependence} shows that, across all three setups, increasing the truncation order brings the lower boundary closer to the Veneziano point.
This trend suggests that the Veneziano amplitude might reach the boundary when the full moment and crossing constraints are imposed.
Panel~(d) shows the remaining gap between the Veneziano amplitude and our numerical results, which is of order $10^{-4}$.
However, as we will see below, this gap does not actually vanish.

\begin{figure}[h]
  \centering
  \begin{minipage}{0.63\textwidth}
    \centering
    \includegraphics[width=\linewidth,height=0.6\textheight,keepaspectratio]{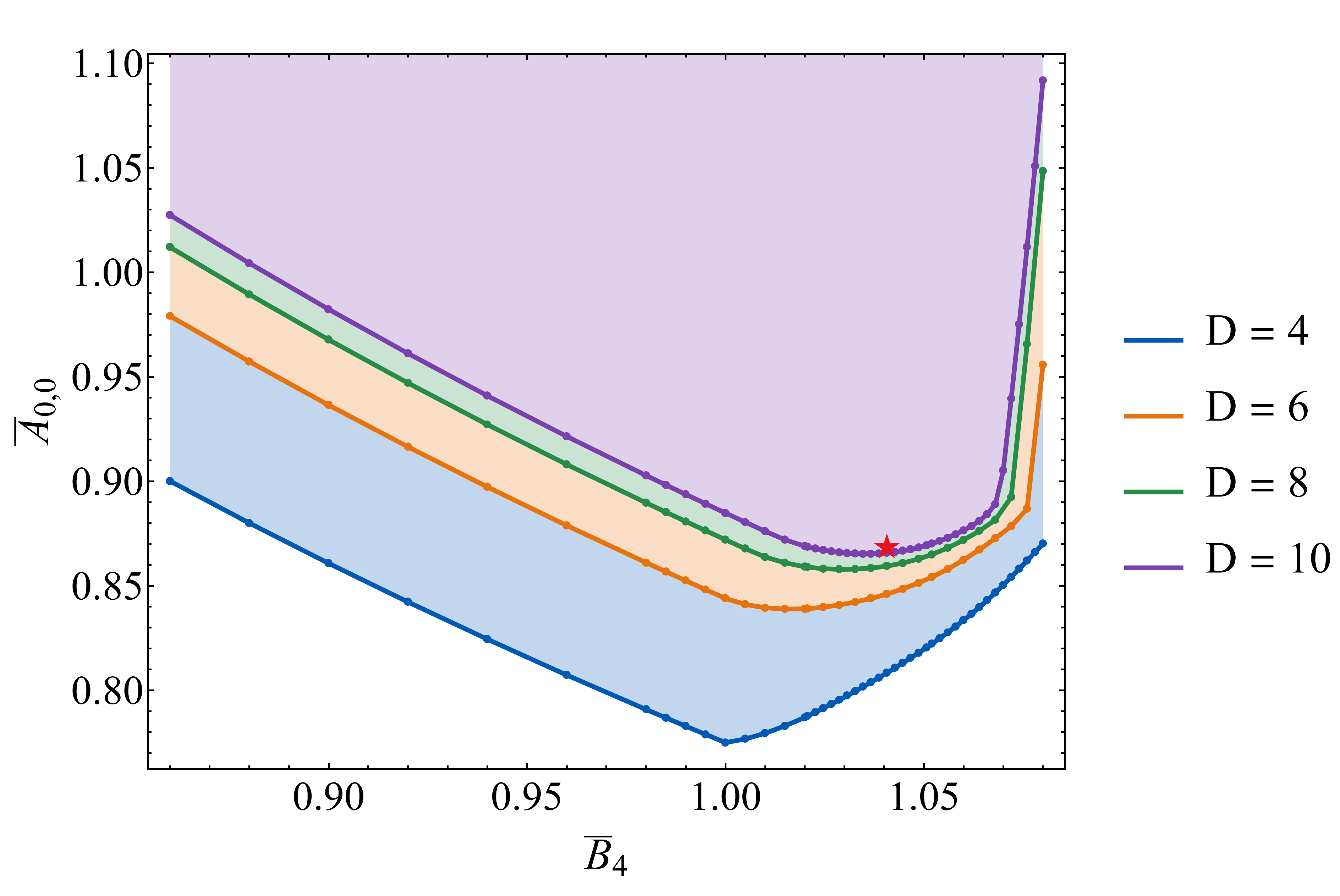}
  \end{minipage}
  \caption{Spacetime-dimension dependence of the numerical bootstrap bounds
  at truncation order $K=21$ in the
  $(\bar{B}_4,\bar{A}_{0,0})$ plane, where
  $\bar{B}_4=B_4/W_{0,1}$ and
  $\bar{A}_{0,0}=A_{0,0}/W_{0,1}$.
  Here $W_{0,1}=7\zeta(4)/4$ is fixed and $A_{1,0}$ is left unconstrained, with no low-lying spectral input. The curves correspond to $D=4,6,8,10$. Shading indicates the regions allowed by the retained constraints. The red star marks the dimension-independent Veneziano point.}
  \label{fig:dimensions}
\end{figure}

Although the Veneziano low-energy data are dimension-independent, the positivity constraints and the resulting boundary geometry depend on $D$.
Figure~\ref{fig:dimensions} shows that the Veneziano point lies further from the boundary of the allowed region as $D$ decreases, whereas the boundary is very close to the Veneziano point in $D=10$.
This is consistent with the increasingly stringent positivity constraints in higher dimensions, with $D=10$ being the limiting dimension for Veneziano partial-wave positivity \cite{Mansfield:2025gca}.

\section{Extremality with fixed leading Wilson data?}
\label{sec:counterexamples}

We now test whether the apparent extremality suggested by the $D=10$ numerical results in Section~\ref{sec:numerical} extends to exact extremality.
We find that this is not the case: there exist counterexamples preserving the prescribed mass gap and Wilson coefficients while lowering $A_{0,0}$.
Throughout this section, we set $\alpha'=1$ and take the EFT cutoff $\Lambda$ entering the dispersion relation to be unity.

In searching for such counterexamples, we naturally start from deformations of the Veneziano amplitude that retain its equally spaced mass-squared spectrum and are related by simple recurrence relations, which will simplify the subsequent positivity proof.
We require each amplitude to satisfy dispersive positivity and then seek counterexamples among their linear combinations, rather than trying to satisfy all the requirements in a single step.
Positive linear combinations automatically preserve dispersive positivity.
To lower $A_{0,0}$ at fixed higher-order coefficients, one might try deformations with more slowly decaying partial-wave coefficients.
Such slower decay, however, can make it difficult to maintain the fixed-${u}$ decay required by the unsubtracted dispersion relation.

This motivates the opposite strategy: introducing small negative contributions from more rapidly decaying deformations.
The remaining task is to ensure partial-wave positivity of the resulting amplitude.
Heuristically, the rapid decay of the deformation's partial-wave coefficients should make positivity easy to maintain at high mass levels, while positivity at the remaining low levels should be preserved by taking the perturbation sufficiently small.
This strategy leads to explicit counterexamples, whose construction we present in detail below.

\subsection{Deformed Veneziano amplitudes}

Our counterexamples use the following deformed Veneziano amplitudes:
\begin{align}
\mathcal A^{(\gamma)}(s,{u})
&=-\frac{1}{s{u}}+\frac{1}{(3\gamma)!}\int_0^1\dd {x}\int_{1-{x}}^1\dd {y}\,\frac{({x}+{y}-1)^{3\gamma}}{{x}^{s+\gamma+1}{y}^{u+\gamma+1}} ,
\end{align}
which are minor variants of the hypergeometric family introduced in Ref~\cite{Cheung:2023adk} and studied in detail in Ref~\cite{Jepsen:2025baw}. 

Let us first explain how this family arises naturally from the strategy outlined above. For the small negative deformation of the Veneziano amplitude, preserving partial-wave positivity of the full amplitude requires its mass--spin support to be contained in that of Veneziano.
Integer shifts provide a natural way to retain the equally spaced mass-squared spectrum, while symmetric combinations preserve crossing.
Additionally, Veneziano spin parity $\ell\equiv N-1\pmod2$ requires the residue reflection condition about $u=-N/2$: $R_N(-N-{u})=(-1)^{N-1}R_N({u})$.
Considering all of these motivates the following deformation
\begin{align}
\mathcal A^{(\gamma)}_{\Gamma}(s,{u})
\equiv\frac{(\mathcal T_s\mathcal T_{u})^{-\gamma}(\mathcal T_s+\mathcal T_{u}-1)^{3\gamma}}{(3\gamma)!}\mathcal A_{\rm V}(s,{u})=(-1)^{\gamma+1}\frac{\Gamma(-\gamma-s)\Gamma(-\gamma-{u})}{\Gamma(1+\gamma-s-{u})},
\end{align}
where the backward-shift operators are defined by
\begin{align}
    \mathcal T_s \mathcal A(s,{u}) \equiv \mathcal A(s-1,{u}), \qquad \mathcal T_{u} \mathcal A(s,{u}) \equiv \mathcal A(s,{u}-1).
\end{align}
However, for $\gamma>0$, the low-lying poles of $\mathcal A^{(\gamma)}_{\Gamma}(s,u)$, located at $-\gamma,\ldots,2\gamma$ in both channels, include tachyonic poles and massive poles with non-polynomial residues.
These low-lying poles must be removed to preserve dispersive positivity to get the desired negative deformation.
A natural way to remove these unwanted poles is to add a crossing-symmetric rational correction that leaves the polynomial residues at $N\geq2\gamma+1$ unchanged, an approach also considered in Ref~\cite{Jepsen:2025baw}.
In fact, the required correction can be obtained simply by applying the same shift operator to the massless term and reversing the sign:
\begin{align}
\mathcal A^{(\gamma)}_{\text{R}}(s,{u})
&\equiv-\frac{(\mathcal T_s\mathcal T_{u})^{-\gamma}(\mathcal T_s+\mathcal T_{u}-1)^{3\gamma}}{(3\gamma)!}\left(-\frac{1}{s{u}}\right).
\label{eq:rational-amplitude}
\end{align}
This construction admits a double Mellin representation that makes the cancellation of the low-lying poles explicit.
To see this,
we start from the regular Veneziano amplitude,
\begin{align}
\widehat{\mathcal A}_{\rm V}(s,u)\equiv\mathcal A_{\rm V}(s,u)+\frac{1}{su}=\int_0^1\dd x\int_{1-x}^1\dd y\,x^{-s-1}y^{-u-1},\qquad\Re s,\Re u<1 .
\end{align}
Note that $\mathcal T_s$ and $\mathcal T_u$ multiply the Mellin integrand by $x$ and $y$, respectively, so the pole-subtracted amplitude is given by
\begin{align}
\widehat{\mathcal A}^{(\gamma)}(s,u)
& =\mathcal A^{(\gamma)}_{\Gamma}(s,{u})+\mathcal A^{(\gamma)}_{\text{R}}(s,{u}) =\frac{1}{(3\gamma)!}\int_0^1\dd x\int_{1-x}^1\dd y\,\frac{(x+y-1)^{3\gamma}}{x^{s+\gamma+1}y^{u+\gamma+1}}.
\end{align}
With $y=1-x+xw$, expanding $[1-x(1-w)]^{-u-\gamma-1}$ and evaluating the beta integral yields
\begin{align}
\widehat{\mathcal A}^{(\gamma)}(s,u) =\sum_{N=2\gamma+1}^{\infty}\frac{R_N^{(\gamma)}(u)}{N-s}, ~~~ R_N^{(\gamma)}(u)=\frac{(u+\gamma+1)_{N-2\gamma-1}}{(N+\gamma)!},~~~ N\geq2\gamma+1,
\end{align}
where we use the Pochhammer symbol $(a)_m\equiv\Gamma(a+m)/\Gamma(a)$.
Thus, the poles at $N\leq2\gamma$ are absent, while the residues at $N\geq2\gamma+1$ remain unchanged.
The retained residues have degree $N-2\gamma-1$ and satisfy $R_N^{(\gamma)}(-N-u)=(-1)^{N-1}R_N^{(\gamma)}(u)$, so their partial-wave expansions contain only spins satisfying
\begin{align}
0\leq\ell\leq N-2\gamma-1,\qquad\ell\equiv N-1\pmod 2.
\end{align}
Therefore, the mass--spin support of these deformation blocks is contained in that of the Veneziano amplitude.

The partial-wave positivity of $\mathcal A^{(\gamma)}(s,u)$ can be established using a simple recurrence relation between successive deformations
\begin{align}
R_N^{(\gamma+1)}({u})=\frac{\Delta_{u}^2R_N^{(\gamma)}({u})}{(N+\gamma+1)(N-2\gamma-2)(N-2\gamma-1)},\qquad N\geq2\gamma+3 ,
\label{eq:residue-recursion}
\end{align}
which follows from the explicit residue formula above and where the central second difference is defined by $\Delta_{u}^2f({u})\equiv f({u}+1)-2f({u})+f({u}-1)$.
Note that the denominator of Eq.~\eqref{eq:residue-recursion} is obviously positive, and $\Delta_{u}^2 f(u)$ also preserves dispersive positivity.
To see the latter, note that the central second difference on polynomials is a finite positive linear combination of even derivatives,
\begin{align}
\Delta_{u}^2f({u})=2\sum_{r\geq1}\frac{\partial_{u}^{2r}f({u})}{(2r)!}.
\end{align}
Furthermore, $\partial_{u}^{2r}f(u)$ preserves dispersive positivity, due to the fact that even derivatives preserve Gegenbauer positivity for $\nu>0$, as follows by iterating the identity
\begin{align}
\partial_{u}^2 C_m^{(\nu)}\!\left(1+\frac{2{u}}{N}\right)=\frac4{N^2}
\sum_{\substack{0\leq\ell\leq m-2\\m-\ell\ {\rm even}}}
(\ell+\nu)(m-\ell)(m+\ell+2\nu)C_\ell^{(\nu)}\!\left(1+\frac{2{u}}{N}\right).
\end{align}
Starting from Veneziano positivity for $4\leq D\leq10$ \cite{Mansfield:2025gca}, induction therefore establishes partial-wave positivity of $R_N^{(\gamma)}$ for every integer $\gamma\geq0$.

Recall that lowering $A_{0,0}$ while preserving the chosen higher-order Wilson coefficients requires some of the deformed amplitudes to enter with negative weights, so the individual dispersive positivity of $\mathcal A^{(\gamma)}(s,{u})$ is not enough.
We also need a uniform bound on their partial-wave coefficients relative to those of Veneziano across all mass levels and spins.
Heuristically, the faster decay of the deformation coefficients, together with the freedom to take arbitrarily small perturbation weights, suggests that positivity of the total amplitude can be maintained.
We give explicit counterexamples below and prove the required uniform bound in Section~\ref{sec:positivity}.

\subsection{Explicit counterexamples}
For each $\mathcal A^{(\gamma)}$, we denote the coefficients in Eq.~\eqref{eq:eft} by $A_{m,n}^{(\gamma)}$ and $W_{p,q}^{(\gamma)}$.
For $\gamma=0,1,2$, the first several coefficients are
\begin{align}
(A_{0,0}^{(0)},A_{1,0}^{(0)},W_{0,1}^{(0)})
&=\left(\zeta(2),\zeta(3),\frac74\zeta(4)\right),\\
(A_{0,0}^{(1)},A_{1,0}^{(1)},W_{0,1}^{(1)})
&=\left(\frac{5}{3}-\zeta(2),\frac{29}{24}-\zeta(3),\frac{85}{24}-\zeta(2)-\frac{7}{4}\zeta(4)\right),
\\
(A_{0,0}^{(2)},A_{1,0}^{(2)},W_{0,1}^{(2)})
&=\left(\frac{\zeta(2)}8-\frac{37}{180},
\frac{\zeta(3)}8-\frac{2077}{13824},
\frac5{32}\zeta(2)+\frac7{32}\zeta(4)-\frac{204779}{414720}\right).
\end{align}
These coefficients can be computed from the pole expansion using the following relations:
\begin{align}
A_{m,n}^{(\gamma)}&=\frac{1}{n!}\!\!\sum_{N\geq2\gamma+1}\!\!\!\frac{\partial_{u}^nR_N^{(\gamma)}(0)}{N^{m-n+1}}, ~ R_N^{(\gamma)}(0)=\frac{(N-\gamma-1)!}{\gamma!(N+\gamma)!},~ \frac{\partial_{u}R_N^{(\gamma)}(0)}{R_N^{(\gamma)}(0)}=H_{N-\gamma-1}-H_\gamma,
\end{align}
where $H_N=\sum_{r=1}^N r^{-1}$ is the $N$th harmonic number, with $H_0=0$.
Computing $A_{0,0}^{(\gamma)}$ and $A_{1,0}^{(\gamma)}$ requires sums of rational functions of $N$, while $W_{0,1}^{(\gamma)}=2A_{2,0}^{(\gamma)}-A_{2,1}^{(\gamma)}$ also involves sums weighted by harmonic numbers.
For the harmonic sums, we first eliminate the shifts in $H_{N-\gamma-1}$ and then apply partial fractions and summation by parts, using $\sum_{N=1}^{\infty}H_N/N^3=\frac54\zeta(4)$.
All resulting sums reduce to zeta values and finite endpoint terms, with individually divergent terms combined at a common upper cutoff before taking the limit.
For example,
\begin{align}
A_{0,0}^{(1)}&=\sum_{N=3}^{\infty}\frac{R_N^{(1)}(0)}{N}=\sum_{N=3}^{\infty}\frac{1}{(N-1)N^2(N+1)}=\sum_{N=3}^{\infty}\left[\frac{1}{2(N-1)}-\frac{1}{N^2}-\frac{1}{2(N+1)}\right]\nonumber\\
&=\lim_{M\to\infty}\left[\frac12\left(\frac12+\frac13-\frac1M-\frac{1}{M+1}\right)-\sum_{N=3}^{M}\frac{1}{N^2}\right]=\frac{5}{12}-\left[\zeta(2)-\frac54\right]=\frac53-\zeta(2).\nonumber
\end{align}

\paragraph{Counterexample with $W_{0,1}=W_{0,1}^{(0)}$.} In this case,
the amplitude
\begin{align}
\mathcal A_\epsilon^{\rm I}(s,{u})
&\equiv-\frac{1}{s{u}}+\frac{\widehat{\mathcal A}^{(0)}(s,{u})-\epsilon\widehat{\mathcal A}^{(2)}(s,{u})}{1-\epsilon W_{0,1}^{(2)}/W_{0,1}^{(0)}}
\end{align}
obeys the unsubtracted fixed-${u}$ dispersion relation and has mass--spin spectrum contained in that of the Veneziano amplitude.
The normalization ensures $W_{0,1}^{\rm I}=W_{0,1}^{(0)}$ while
\begin{align}
A_{0,0}^{\rm I}-\zeta(2)
&=\epsilon \frac{W_{0,1}^{(2)} A_{0,0}^{(0)}-W_{0,1}^{(0)}A_{0,0}^{(2)}}
{W_{0,1}^{(0)}-\epsilon W_{0,1}^{(2)}},
\end{align}
where the numerator is approximately $-1.12\times10^{-4}$, while the denominator is positive throughout $0 < \epsilon \leq 1/1024$, since $W_{0,1}^{(0)}\approx1.89$ and $W_{0,1}^{(2)}\approx2.63\times10^{-6}$.
Hence $A_{0,0}^{\rm I}<\zeta(2)$ throughout this interval.

\paragraph{Counterexample with $A_{1,0}=A_{1,0}^{(0)}$ and $W_{0,1}=W_{0,1}^{(0)}$.}
To preserve $A_{1,0}=\zeta(3)$ as well, consider the amplitude
\begin{align}
\mathcal A_\epsilon^{\rm II}(s,{u})
&\equiv-\frac{1}{s{u}}+(1-\kappa\epsilon)\widehat{\mathcal A}^{(0)} (s,{u})
+\lambda\epsilon\widehat{\mathcal A}^{(1)} (s,{u})-\epsilon\widehat{\mathcal A}^{(2)} (s,{u}).
\end{align}
This amplitude obeys the unsubtracted fixed-${u}$ dispersion relation and retains the unit mass gap and equally spaced mass-squared spectrum.
Choosing $\lambda$ and $\kappa$ to satisfy
\begin{align}
-\kappa\zeta(3)+\lambda A_{1,0}^{(1)}=A_{1,0}^{(2)},~~~~~~~~~~~~
-\kappa W_{0,1}^{(0)}+\lambda W_{0,1}^{(1)}=W_{0,1}^{(2)}
\end{align}
ensures $A_{1,0}^{\rm II}=\zeta(3)$ and $W_{0,1}^{\rm II}=W_{0,1}^{(0)}$.
The unique solution has $\lambda\approx2.07\times10^{-3}$ and $\kappa\approx1.53\times10^{-6}$, while
\begin{align}
A_{0,0}^{\rm II}-\zeta(2)
&=\epsilon[-\kappa\zeta(2)+\lambda A_{0,0}^{(1)}-A_{0,0}^{(2)}]\approx-1.87\times10^{-5}\epsilon.
\end{align}
Thus $A_{0,0}^{\rm II}<\zeta(2)$ for $\epsilon>0$.

In the next subsection, we will first prove dispersive positivity for both examples in $D=10$ on $0<\epsilon\leq1/1024$\footnote{The value $1/1024$ is chosen for convenience in this analytic proof and is not intended to be optimal.
Our preliminary numerical tests suggest upper endpoints of order $\sim 0.1$ for both examples.
Establishing positivity for larger $\epsilon$ requires sharper analytic estimates, while locating the endpoints reliably requires high-precision numerics; we therefore quote this scale only as a rough indication.}, focusing on the uniform bound needed for the first; the second follows from the same bound and the positive $\gamma=1$ contribution.
The extension to $4\leq D\leq10$ follows directly from the Gegenbauer connection formula \cite[Theorem~7.1.4$'$]{andrews1999special}, valid for $0<\nu'\leq\nu$:
\begin{align}
C_J^{(\nu)}(x)&=\sum_{q=0}^{\lfloor J/2\rfloor}\frac{(\nu-\nu')_q(\nu)_{J-q}(J-2q+\nu')}{\nu' q!(\nu'+1)_{J-q}}C_{J-2q}^{(\nu')}(x).
\label{eq:gegenbauer}
\end{align}
The coefficients remain non-negative after endpoint normalization.
Taking $\nu=7/2$ and $\nu'=(D-3)/2$ therefore preserves both positivity and the coefficientwise bounds established in $D=10$.
The construction also suggests that further blocks might preserve more general finite sets of higher-order Wilson data while lowering $A_{0,0}$, although this extension is beyond the scope of the current paper.

\subsection{All-level positivity}
\label{sec:positivity}

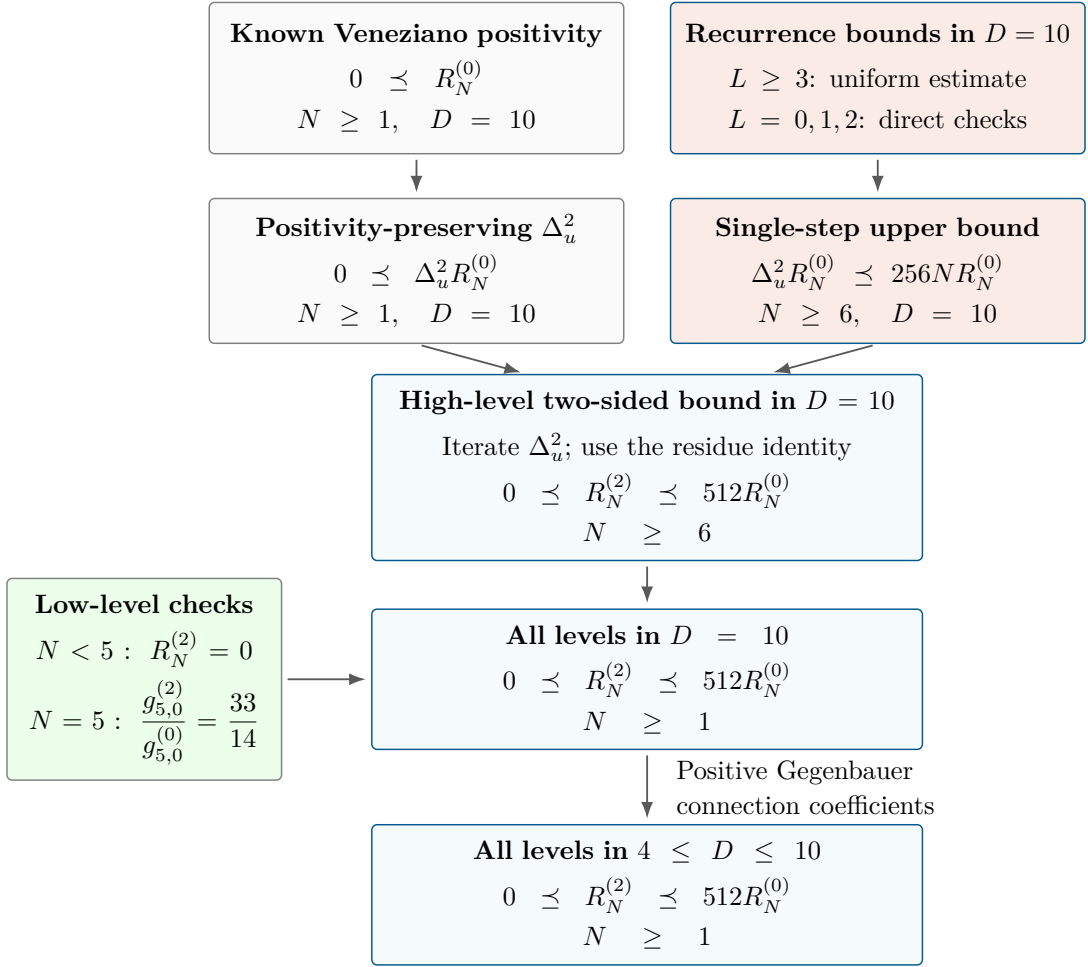
\begin{figure}[!htbp]
\centering
\begin{tikzpicture}[
  x=1cm,y=1cm,
  font=\small,
  proofbox/.style={draw=black!50,line width=0.55pt,rounded corners=2pt,fill=black!2,align=center,text width=5.2cm,inner xsep=4pt,inner ysep=6pt,minimum height=1.75cm},
  estimate/.style={proofbox,draw=contourblue!75!black,fill=contourblue!4},
  proofarrow/.style={-{Latex[length=2.2mm,width=1.6mm]},draw=black!65,line width=0.7pt,shorten <=2pt,shorten >=2pt}
]
\node[proofbox,minimum height=2cm] (input) at (-3.05,0)
  {\textbf{Known Veneziano positivity}\\[4pt]
  $0\preceq R_N^{(0)}$\\[2pt]
  $N\geq1,\quad D=10$};
\node[estimate,minimum height=2cm,fill=cutorange!10] (recurrence) at (3.05,0)
  {\textbf{Recurrence bounds in $D=10$}\\[4pt]
  $L\geq3$: uniform estimate\\[2pt]
  $L=0,1,2$: direct checks};

\node[proofbox] (positive) at (-3.05,-2.55)
  {\textbf{Positivity-preserving $\Delta_u^2$}\\[4pt]
  $0\preceq\Delta_u^2R_N^{(0)}$\\[2pt]
  $N\geq1,\quad D=10$};
\node[estimate,fill=cutorange!10] (single) at (3.05,-2.55)
  {\textbf{Single-step upper bound}\vphantom{$\Delta_u^2$}\\[4pt]
  $\Delta_u^2R_N^{(0)}\preceq256N R_N^{(0)}$\\[2pt]
  $N\geq6,\quad D=10$};

\node[estimate,text width=7cm,minimum height=2.1cm] (large) at (0,-5.15)
  {\textbf{High-level two-sided bound in $D=10$}\\[4pt]
  Iterate $\Delta_u^2$; use the residue identity\\[4pt]
  $0\preceq R_N^{(2)}\preceq512R_N^{(0)}$\\[2pt]
  $N\geq6$};
\node[proofbox,text width=3.35cm,minimum height=1.9cm,fill=green!8!white] (low) at (-6.65,-7.95)
  {\textbf{Low-level checks}\\[4pt]
  $N<5:\ R_N^{(2)}=0$\\[4pt]
  $N=5:\ \dfrac{g_{5,0}^{(2)}}{g_{5,0}^{(0)}}=\dfrac{33}{14}$};
\node[estimate,text width=7cm] (all) at (0,-7.95)
  {\textbf{All levels in $D=10$}\\[4pt]
  $0\preceq R_N^{(2)}\preceq512R_N^{(0)}$\\[2pt]
  $N\geq1$};
\node[estimate,text width=7cm] (dimensions) at (0,-10.8)
  {\textbf{All levels in $4\leq D\leq10$}\\[4pt]
  $0\preceq R_N^{(2)}\preceq512R_N^{(0)}$\\[2pt]
  $N\geq1$};

\draw[proofarrow] (input.south) -- (positive.north);
\draw[proofarrow] (recurrence.south) -- (single.north);
\draw[proofarrow] (positive.south) -- ([xshift=-1.6cm]large.north);
\draw[proofarrow] (single.south) -- ([xshift=1.6cm]large.north);
\draw[proofarrow] (large.south) -- (all.north);
\draw[proofarrow] (low.east) -- (all.west);
\draw[proofarrow] (all.south) -- (dimensions.north)
  node[midway,right=7pt,align=left,font=\small] {Positive Gegenbauer\\connection coefficients};
\end{tikzpicture}
\caption{Outline of the proof of Eq.~\eqref{eq:residue-bound}. $g_{N,\ell}^{(\gamma)}$ denotes the coefficient of $\overline C_\ell^{(\nu)}(1+2{u}/N)$ in $R_N^{(\gamma)}({u})$ and $N=\ell+2L+1$.}
\label{fig:positivity-flow}
\end{figure}

To prove positivity of both counterexamples in the last subsection for $4\leq D\leq10$, the key is to establish the following sufficient uniform bound:
\begin{align}
R_N^{(2)}\preceq512R_N^{(0)},\qquad N\geq1,
\label{eq:residue-bound}
\end{align}
where we have used the shorthand $f\preceq g$ if $g-f$ has non-negative coefficients in the basis $\overline C_\ell^{(\nu)}(1+2{u}/N)$ at fixed mass level $N$.
The upper bound $R_N^{(2)}\preceq512R_N^{(0)}$ for $N\geq6$ is the more difficult one to establish, which follows if we can show
\begin{align}
\Delta_{u}^2R_N^{(0)}\preceq256NR_N^{(0)} .
\label{eq:difference-bound}
\end{align}
If the latter is established, we have $(\Delta_{u}^2)^2R_N^{(0)}\preceq(256N)^2R_N^{(0)}$.
Together with the identity $(\Delta_{u}^2)^2R_N^{(0)}
=(N^2-1)(N^2-4)(N-3)(N-4)R_N^{(2)}$ $(N\geq5)$, we can get the desired upper bound $R_N^{(2)}\preceq512R_N^{(0)}$ for $N\geq 6$.
The low-level $1\leq N< 6$ cases can be easily verified separately.
Our proof builds on the work of Refs~\cite{Chen:2026zmz, Arkani-Hamed:2022gsa, Mansfield:2025gca}.
Figure~\ref{fig:positivity-flow} summarizes the main steps of the proof.

First, note that the low levels can be settled immediately.
For $N<5$, $R_N^{(2)}=0$ and Veneziano positivity gives Eq.~\eqref{eq:residue-bound}.
Let $g_{N,\ell}^{(\gamma)}$ denote the coefficient of $\overline C_\ell^{(\nu)}(1+2{u}/N)$ in the expansion of $R_N^{(\gamma)}({u})$.
At $N=5$, the $\gamma=2$ residue is a scalar, and direct projection in $D=10$ gives
\begin{align}
g_{5,0}^{(0)}=\frac{1}{11880},\qquad g_{5,0}^{(2)}=\frac{1}{5040},\qquad \frac{g_{5,0}^{(2)}}{g_{5,0}^{(0)}}=\frac{33}{14}<512 .
\end{align}
All higher-spin coefficients of $R_5^{(2)}$ vanish, so $R_5^{(2)}\preceq512R_5^{(0)}$.

For $N\geq6$, as mentioned, we must establish the upper bound Eq.~\eqref{eq:difference-bound}, which can be obtained by comparing the partial-wave coefficients of $\Delta_u^2R_N^{(0)}$ with those of $R_N^{(0)}$.
Both sets of coefficients contain the positive prefactor
\begin{align}
\mathcal N_{N,\ell}^{(0)}\equiv\frac{2^{1-N-\ell}N^\ell C_\ell^{(\nu)}(1)}{N(\nu)_\ell}>0.
\end{align}
We use either $(N,\ell)$ or $(L,\ell)$ to label the same coefficients, with $L=(N-\ell-1)/2$ being more convenient for the recurrence below.
For both polynomials, coefficients with $L\notin\mathbb Z_{\geq0}$ vanish.
Let us write the Veneziano partial-wave coefficient as
\begin{align}
g_{N,\ell}^{(0)}=\mathcal N_{N,\ell}^{(0)}h_L(L,\ell),
\end{align}
and then the corresponding partial-wave coefficient of $\Delta_u^2R_N^{(0)}$ is $4\mathcal N_{N,\ell}^{(0)}\sum_{r=1}^Lv_rh_{L-r}(L,\ell)$, where $v_r=2^{2r-1}/(2r)!$. $h_p(L,\ell)$ can be determined recursively from $h_0(L,\ell)=1$:
\begin{align}
h_p(L,\ell)&=\sum_{r=2}^{p+1}C_{p,r}(L,\ell)h_{p+1-r}(L,\ell),\qquad p\geq1,
\label{eq:h-recursion}\\
C_{p,r}(L,\ell)&=\frac{v_rD_{p,r}(L,\ell)}{4p(p+\ell+\nu)},
\\
D_{p,r}(L,\ell)&=N\left[(4L+1-2\nu)r+\ell-2L+2\nu-1\right]\nonumber\\
&\quad-2(p+1-r)\left[2(p+1-r)+2Nr+2\nu-4L-2\right].
\end{align}
To derive this formula, we can first evaluate the Gegenbauer projection using a contour representation of the residues, and
then match powers in the differential equation for the corresponding generating function; see Appendix~\ref{sec:appendix-projection} for details.

We now specialize to $\nu=7/2$ ($D=10$) for the uniform bound in Eq.~\eqref{eq:difference-bound}.
After cancelling a positive prefactor, it suffices to prove the following coefficientwise bound:
\begin{align}
S_L(L,\ell)\equiv\sum_{r=1}^Lv_rh_{L-r}(L,\ell)\leq64N h_L(L,\ell),\qquad N\geq6.
\label{eq:coefficient-bound}
\end{align}
The validity of this bound for asymptotic $\ell$ is easy to establish.
To see this, for fixed $L,p,r$, we have $C_{p,r}(L,\ell)\simeq {v_r\ell}/{4p}$ as $\ell\to \infty$.
Thus, the recurrence relation in Eq.~\eqref{eq:h-recursion} implies that the leading power of $\ell$ in $h_p$ is given by
\begin{align}
h_p(L,\ell)&=\frac{\ell}{12p}h_{p-1}(L,\ell)+O_{L,p}(\ell^{p-1})=\frac{\ell^p}{12^p p!}+O_{L,p}(\ell^{p-1}),\qquad1\leq p\leq L.
\end{align}
Since $v_1=1$, $S_L=\sum_{r=1}^Lv_rh_{L-r}$ is dominated by its $r=1$ term:
\begin{align}
\frac{S_L(L,\ell)}{h_L(L,\ell)}&=\frac{12L}{\ell}+O_L(\ell^{-2}),\qquad L\geq1.
\end{align}
Thus the bound in Eq.~\eqref{eq:coefficient-bound} is proven for fixed $L$ and large $\ell$.
The full proof of the uniform bound in Eq.~\eqref{eq:coefficient-bound} away from large $\ell$ is given in Appendix~\ref{sec:positivity-details}.
Together with the low-level checks above, the uniform bound implies Eq.~\eqref{eq:residue-bound} in $D=10$, or equivalently
\begin{align}
0\leq g_{N,\ell}^{(2)}\leq512g_{N,\ell}^{(0)},\qquad N\geq1,\quad\ell\geq0.
\label{eq:partial-wave-bound}
\end{align}

Let $g_{N,\ell}^{\rm I}$ and $g_{N,\ell}^{\rm II}$ denote the partial-wave coefficients of counterexamples $\mathcal A_\epsilon^{\rm I}$ and $\mathcal A_\epsilon^{\rm II}$ at mass level $N$, respectively.
For $0<\epsilon\leq1/1024$, Eq.~\eqref{eq:partial-wave-bound} and positivity of the $\gamma=1$ contribution give, in $D=10$,
\begin{align}
g_{N,\ell}^{\rm I}&=\frac{g_{N,\ell}^{(0)}-\epsilon g_{N,\ell}^{(2)}}{1-\epsilon W_{0,1}^{(2)}/W_{0,1}^{(0)}}\geq\frac{(1-512\epsilon)g_{N,\ell}^{(0)}}{1-\epsilon W_{0,1}^{(2)}/W_{0,1}^{(0)}}\geq0,\\
g_{N,\ell}^{\rm II}&=(1-\kappa\epsilon)g_{N,\ell}^{(0)}+\lambda\epsilon g_{N,\ell}^{(1)}-\epsilon g_{N,\ell}^{(2)}\geq[1-(512+\kappa)\epsilon]g_{N,\ell}^{(0)}+\lambda\epsilon g_{N,\ell}^{(1)}\geq0.
\end{align}
Here $\kappa\approx1.53\times10^{-6}$ and $\lambda\approx2.07\times10^{-3}$.
The non-negative Gegenbauer connection coefficients in Eq.~\eqref{eq:gegenbauer} then extend partial-wave positivity of both amplitudes to $4\leq D\leq10$.

This completes the counterexamples: the prescribed one- and two-coefficient constraints do not make Veneziano an exact minimizer, despite the close numerical bounds.

\section{Extremality under forward ultrasoftness}
\label{sec:analytic}

The counterexamples of Section~\ref{sec:counterexamples} rule out exact Veneziano extremality from $A_{0,0}$ by simply specifying some leading Wilson data.
In this section, we shall explore the $A_{0,0}$  extremality by alternatively imposing a UV softness condition motivated by the Kawai--Lewellen--Tye (KLT) relation \cite{Kawai:1985xq}.
This UV condition controls the amplitude near the complex forward line $u=-s$, and, within the dispersive bootstrap framework, allows us to express $A_{0,0}-(\alpha')^2\zeta(2)$ as a non-negative spectral integral.
This gives the sharp bound $A_{0,0}\geq(\alpha')^2\zeta(2)$ at fixed $\alpha'$.
The resulting spectral functional is analogous to extremal functionals in the conformal bootstrap \cite{Mazac:2018mdx, Mazac:2018ycv}.
At saturation, the zeros of its kernel enforce the equally spaced spectrum and the corresponding open-string spin parity, which in turn uniquely selects the Veneziano amplitude for $4\leq D\leq10$.

\subsection{KLT and forward ultrasoftness}

The Kawai--Lewellen--Tye (KLT) relation \cite{Kawai:1985xq} motivates the UV behavior relevant to our extremal problem.
Suppressing couplings and polarization-dependent kinematic factors, this relation connects four-point closed and open string amplitudes and takes the schematic form
\begin{align}
\mathcal M_{\rm KLT}(s,t,u)\propto\sin(\pi\alpha's)\mathcal A(s,t)\mathcal A(s,u),\qquad u=-s-t,
\end{align}
where $\alpha'$ is the open-string Regge slope.
For $s=Re^{\ii\theta}$ with fixed $\theta\notin\pi\mathbb Z$,
$2|\sin(\pi\alpha's)|\sim e^{\pi\alpha'R|\sin\theta|}
$ as $R\to\infty$.
Also, at fixed $t$, the stripped amplitudes $\mathcal M_{\rm KLT}(s,t,u)$ and $\mathcal A(s,t)$ are expected to be bounded polynomially at large $s$.
Thus, the crossed amplitude $\mathcal A(s,u)$ is expected to be exponentially suppressed.
Guided by this KLT argument, we impose forward ultrasoftness by requiring, for fixed real $t\leq0$ near zero,
\begin{align}
e^{\ii\pi\alpha's}\mathcal A(s,-s-t)\to0,\qquad |s|\to\infty,
\label{eq:ultrasoftness}
\end{align}
along large contours that avoid the poles.
This condition excludes the counterexamples of Section~\ref{sec:counterexamples}.\footnote{This is because the rational terms in Eq.~\eqref{eq:rational-amplitude} $\widehat{\mathcal A}^{(\gamma)}(s,-s)=-{1}/[{(3\gamma)!s^2}]+O(s^{-4})$ decay polynomially, and thus violate the forward ultrasoftness.}
We now use this UV condition to derive a sharp lower bound on $A_{0,0}$.

\subsection{Lower bound on $A_{0,0}$}

\begin{figure}[t]
\centering
\resizebox{0.7\linewidth}{!}{
\begin{tikzpicture}[
  x=1cm,y=1cm,
  font=\scriptsize,
  >=Latex,
  axis/.style={draw=black!55,line width=0.4pt,-{Latex[length=1.6mm]}},
  cut/.style={draw=cutorange,line width=1.65pt,line cap=round},
  contour/.style={draw=contourblue,line width=0.95pt,postaction={decorate},decoration={markings,mark=at position 0.55 with {\arrow{>}}}},
  outer/.style={contour,dash pattern=on 3.2pt off 2.2pt}
]
\pgfmathsetmacro{\keyholeHeight}{0.24}
\pgfmathsetmacro{\keyholeAngle}{asin(\keyholeHeight/2.25)}
\pgfmathsetmacro{\keyholeX}{3.5*cos(\keyholeAngle)}

\draw[axis] (-4.15,0) -- (4.15,0) node[right,font=\footnotesize] {$\operatorname{Re}s$};
\draw[axis] (0,-2.75) -- (0,2.95) node[above,font=\footnotesize] {$\operatorname{Im}s$};
\draw[cut] (-4,0) -- (-1,0);
\draw[cut] (1,0) -- (4,0);

\draw[outer] (\keyholeX,\keyholeHeight) arc[start angle=\keyholeAngle,end angle={180-\keyholeAngle},x radius=3.5cm,y radius=2.25cm];
\draw[outer] (-\keyholeX,-\keyholeHeight) arc[start angle={180+\keyholeAngle},end angle={360-\keyholeAngle},x radius=3.5cm,y radius=2.25cm];
\draw[contour] (-\keyholeX,\keyholeHeight) -- (-1,\keyholeHeight);
\draw[contour] (-1,-\keyholeHeight) -- (-\keyholeX,-\keyholeHeight);
\draw[contour] (\keyholeX,-\keyholeHeight) -- (1,-\keyholeHeight);
\draw[contour] (1,\keyholeHeight) -- (\keyholeX,\keyholeHeight);
\draw[contourblue,line width=0.95pt] (-1,\keyholeHeight) arc[start angle=90,end angle=-90,radius=0.24cm];
\draw[contourblue,line width=0.95pt] (1,-\keyholeHeight) arc[start angle=-90,end angle=-270,radius=0.24cm];

\fill[cutorange] (-1,0) circle [radius=1.2pt];
\fill[cutorange] (1,0) circle [radius=1.2pt];
\draw[fill=white,line width=0.65pt] (0,0) circle [radius=2.4pt];
\fill (0,0) circle [radius=0.65pt];
\node[below right=4pt] at (0,0) {$0$};

\node[text=contourblue] at (0.95,2.56) {$\mathcal C_{\circ}(R)$};
\node[text=contourblue] at (-2.25,0.7) {$\mathcal C_{\rm cut}(R)$};
\node[text=contourblue] at (2.25,0.7) {$\mathcal C_{\rm cut}(R)$};
\node[text=cutorange] at (-2.35,-0.72) {$(-\infty,-\Lambda^2]$};
\node[text=cutorange] at (2.35,-0.72) {$[\Lambda^2,\infty)$};
\node at (-1,-0.7) {$-\Lambda^2$};
\node at (1,-0.7) {$\Lambda^2$};
\end{tikzpicture}
}
\caption{The keyhole contour $\mathcal C(R)=\mathcal C_{\circ}(R)\cup\mathcal C_{\rm cut}(R)$ for the forward restriction $\mathcal A(s,-s)$.
The forward ultrasoftness assumption removes the dashed outer component $\mathcal C_{\circ}(R)$, while $\mathcal C_{\rm cut}(R)$ comprises the two dogbones around the cuts.}
\label{fig:keyhole}
\end{figure}
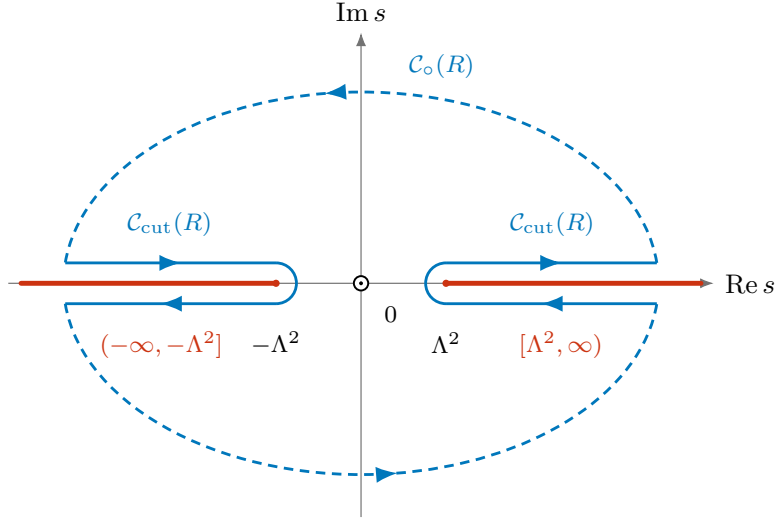

Now, we prove $A_{0,0} \geq (\alpha')^2\zeta(2)$ at fixed $\alpha'$.
To derive this bound, we introduce the normalized forward function
\begin{align}
\mathcal G_\beta(s)\equiv\frac{\mathcal A(s,-s)}{\mathcal A_{\rm V}^{\beta}(s,-s)}=\frac{s\sin(\pi\beta s)}{\pi\beta}\mathcal A(s,-s),\qquad \mathcal G_\beta(0)=1.
\end{align}
At $\beta=\alpha'$, $\mathcal G_{\alpha'}=1$ for the Veneziano amplitude.
We extract its quadratic coefficient with the contour functional
\begin{align}
\mathcal I(\beta;R)\equiv\frac{1}{2\pi\ii}\oint_{\mathcal C(R)} \mathcal G_\beta(s)\frac{\dd s}{s^3},\qquad 0<\beta\leq\alpha' ,
\end{align}
where $\beta$ is an auxiliary scale and $\mathcal C(R)$ is the keyhole contour shown in Figure~\ref{fig:keyhole}.
The low-energy expansion gives $\mathcal G_\beta(s)=1+\left[A_{0,0}-\beta^2\zeta(2)\right]s^2+O(s^4)$.
The residue theorem therefore gives
\begin{align}
\mathcal I(\beta;R)=A_{0,0}-\beta^2\zeta(2).
\label{eq:sine-residue}
\end{align}
At $\beta=\alpha'$, this is precisely the difference from the Veneziano value.
To express this difference in terms of the spectral data, we first decompose the contour integral as
\begin{align}
\mathcal I(\beta;R)=\mathcal I_{\circ}(\beta;R)+\mathcal I_{\rm cut}(\beta;R),
\end{align}
where $\mathcal I_{\circ}(\beta;R)$ is the outer-contour contribution and $\mathcal I_{\rm cut}(\beta;R)$ is the combined contribution from the two cuts.

Under forward ultrasoftness, the outer-contour contribution $\mathcal I_{\circ}(\beta;R)$ vanishes as $R\to\infty$.
To see this, we choose symmetric outer contours with length $L_R=O(R)$ and $|s|\geq cR$ for some $c>0$.
Eq.~\eqref{eq:ultrasoftness} gives $\epsilon_R\equiv\sup_{s\in\mathcal C_{\circ}(R)}|e^{\ii\pi\alpha's}\mathcal A(s,-s)|\to0$.
Crossing supplies the same bound for the opposite phase, so $|\sin(\pi\beta s)\mathcal A(s,-s)|\leq\epsilon_R$ for $0<\beta\leq\alpha'$.
Consequently, $|\mathcal I_{\circ}(\beta;R)|\leq {L_R\epsilon_R}/({2\pi^2\beta(cR)^2})\to 0$.
Now, only the cut contribution $\mathcal I_{\rm cut}(\beta;R)$ remains.
On the positive cut, the partial-wave argument is $-1$, giving the factor $(-1)^\ell$.
Crossing symmetry makes the two cut contributions equal.
Taking $R\to\infty$ in Eq.~\eqref{eq:sine-residue} thus yields
\begin{align}
\mathcal I(\beta)\equiv\lim_{R\to\infty}\mathcal I(\beta;R)&=A_{0,0}-\beta^2\zeta(2)=2\sum_{\ell=0}^{\infty}\int_{\Lambda^2}^{\infty}\frac{(-1)^\ell\sin(\pi\beta s')}{\pi\beta(s')^2}\rho_\ell(s')\,\dd s'.
\end{align}
To make positivity manifest, we differentiate $\beta\mathcal I(\beta)$ for $0<\beta<\alpha'$,
\begin{align}
\frac{\dd}{\dd\beta}[\beta\mathcal I(\beta)]&=A_{0,0}-3\beta^2\zeta(2)=2\sum_{\ell=0}^{\infty}\int_{\Lambda^2}^{\infty}\frac{(-1)^\ell\cos(\pi\beta s')}{s'}\rho_\ell(s')\,\dd s'.
\label{eq:sine-derivative}
\end{align}
Taking $\beta\to\alpha'$, we combine Eqs.~\eqref{eq:a00} and \eqref{eq:sine-derivative} with coefficients $2/3$ and $1/3$, respectively, to obtain
\begin{align}
\boxed{A_{0,0}-(\alpha')^2\zeta(2)=\frac43\sum_{\ell=0}^{\infty}\int_{\Lambda^2}^{\infty}K_\ell(s')\rho_\ell(s')\,\dd s'\geq0}~,
\label{eq:positive-bound}
\end{align}
where $K_\ell(s')\equiv[1+(-1)^\ell\cos(\pi\alpha's')]/(2s')\geq0$.
The Veneziano amplitude satisfies forward ultrasoftness and is allowed in our bootstrap setup for $4\leq D\leq10$.
Since $A_{0,0}^{\rm V}=(\alpha')^2\zeta(2)$, it saturates the bound.
Note that the kernel emerges from the scale derivative of the KLT-motivated sine sum rule, realizing an extremal functional analogous to those of the conformal bootstrap \cite{Mazac:2018mdx,Mazac:2018ycv}.

\subsection{Uniqueness at saturation}

Saturation of Eq.~\eqref{eq:positive-bound} requires the spectral measures to be supported on the zero set of $K_\ell(s')$:
\begin{align}
s'=s_N\equiv\frac{N}{\alpha'},\qquad \ell\equiv N-1\pmod2,\qquad N\in\mathbb N_+.
\end{align}
This fixes the allowed masses and spin parity, but leaves the spectral weights undetermined.

We now show that the spectral weights coincide with those of the Veneziano amplitude.
The key observation is that saturation selects the open-string spin parity.
At each occupied level $s=s_N$, Gegenbauer parity gives the reflection identity for the residue $R_N(u)$: $R_N(-s_N-u)=(-1)^{N-1}R_N(u)$.
For fixed $t\in(-\epsilon,0)$ with $0<\epsilon<1/\alpha'$, consider the monodromy combination
\begin{align}
    \mathcal S_t(s)\equiv\mathcal A(s,t)+e^{\ii\pi\alpha's}\mathcal A(s,u)+e^{-\ii\pi\alpha't}\mathcal A(t,u).
\end{align}
The phase factors and the residue reflection identity cancel all finite poles of $\mathcal S_t(s)$, making it entire.
The ordinary fixed-$u$ falloff and forward ultrasoftness fix it to zero near $s=0$.
Since $\mathcal S_t(s)$ is entire, this identity holds throughout the complex $s$ plane.
Analytic continuation in $t$ then establishes the monodromy relation throughout the Mandelstam plane $s+t+u=0$: $\mathcal S_t(s)=0$.
The bootstrap result of Ref~\cite{Wan:2026pjq} then implies that the amplitude must be Veneziano
\begin{align}
\mathcal A(s,u)=\mathcal A_{\rm V}^{\alpha'}(s,u)=-(\alpha')^2\frac{\Gamma(-\alpha's)\Gamma(-\alpha'u)}{\Gamma(1-\alpha'(s+u))}.
\end{align}
Therefore, the Veneziano amplitude is the unique minimizer at fixed $\alpha'$ for $4\leq D\leq10$.

\subsection{Extremality and zero-pole balance}
\label{sec:minimal-zeros}

Finally, let us comment that our forward ultrasoftness assumption is similar in spirit to the Regge ultrasoftness assumed in Ref~\cite{Cheung:2025tbr}, where the amplitude exhibits arbitrarily fast power-law decay at increasingly negative momentum transfer.
The minimal-zero requirement of that work, which excludes residue zeros beyond those required by Regge consistency, also parallels our minimization of $A_{0,0}$ as a principle for selecting the string amplitude.
These similarities naturally raise the question of a more direct relation between the Wilson coefficient and the amplitude's zero-pole structure. Indeed, under the additional assumptions stated below, we can relate $A_{0,0}-\zeta(2)$ to the zeros and poles of the normalized forward amplitude.

Setting $\alpha'=1$, we write the normalized forward amplitude as
\begin{align}
  \mathcal G(s)\equiv\mathcal G_1(s)=\frac{\mathcal A(s,-s)}{\mathcal A_{\rm V}^{1}(s,-s)}=\frac{s\sin(\pi s)}{\pi}\mathcal A(s,-s),\qquad \mathcal G(0)=1.
  \label{eq:GsDef1}
\end{align}
For Veneziano, $\mathcal G=1$, so the zeros and poles of $\mathcal G$ describe departures from the Veneziano analytic structure.
Crossing makes $\mathcal G$ even, and its low-energy expansion gives
\begin{align}
  \mathcal G(s)&=1+[A_{0,0}-\zeta(2)]s^2+O(s^4),\qquad A_{0,0}-\zeta(2)=\frac12\partial_s^2\log \mathcal G|_{s\to 0}.
\label{eq:forward-curvature}
\end{align}
To express $A_{0,0}-\zeta(2)$ in terms of the zeros and poles of $\mathcal G$, we require the following sums over its zeros and poles to be finite:
\begin{align}
\sum_{z\in Z(\mathcal G)}\frac{1}{|z|^2}+\sum_{p\in P(\mathcal G)}\frac{1}{|p|^2}<\infty ,
\end{align}
where $Z(\mathcal G)$ and $P(\mathcal G)$ denote all zeros and poles of $\mathcal G$, respectively, with each zero $z$ counted $m_z$ times and each pole $p$ counted $m_p$ times.
There is no zero or pole at the origin because $\mathcal G(0)=1$.
This condition ensures absolute and locally uniform convergence of the following zero and pole products:
\begin{align}
\mathcal Z(s)=\prod_{z\in Z(\mathcal G)}\sqrt{1-\frac{s^2}{z^2}},\qquad \mathcal P(s)=\prod_{p\in P(\mathcal G)}\sqrt{1-\frac{s^2}{p^2}}.
\label{eq:zero-and-pole-product}
\end{align}
Opposite points have equal multiplicities by evenness of $\mathcal G$; each pair of factors is multiplied together before forming the infinite product.
Both products are even entire functions normalized to one at the origin.
$\mathcal Z(s)$ and $\mathcal P(s)$ determine $\mathcal G(s)$ up to a zero-free entire factor:
\begin{align}
  \mathcal G(s)=e^{g(s)}\frac{\mathcal Z(s)}{\mathcal P(s)},\qquad g(0)=0 ,
\end{align}
where $g$ is entire and even. However, the summability condition above, evenness of $\mathcal G$, and a uniform closed-contour version of forward ultrasoftness together imply $g=0$.
The precise proof is given in Appendix~\ref{sec:forward-reconstruction}.
Consequently,
\begin{align}
\mathcal G(s)=\frac{\mathcal Z(s)}{\mathcal P(s)}.
\end{align}
Differentiating the zero and pole products near the origin and using Eq.~\eqref{eq:forward-curvature} now gives
\begin{align}
  A_{0,0}-\zeta(2)=\frac12\sum_{p\in P(\mathcal G)}\frac{1}{p^2}-\frac12\sum_{z\in Z(\mathcal G)}\frac{1}{z^2} \geq 0.
\end{align}
These sums retain the complex phases of $p^{-2}$ and $z^{-2}$, while the earlier summability condition ensures their absolute convergence. At saturation of the $s^2$ coefficient $ A_{0,0}-\zeta(2)=0$, the contributions from the zeros and poles must exactly cancel out. Then, the uniqueness result just established forces all the higher order $s^{2n}$ coefficients in Eq.~\eqref{eq:forward-curvature} to vanish and fixes
\begin{align}
\mathcal G(s)=1,
\end{align}
meaning that the full amplitude is Veneziano.
The full extremal constraints therefore strengthen saturation into matching of the entire zero and pole distributions of the full amplitude.

\section{Summary}
\label{sec:summary}

We have investigated whether minimizing the leading $F^4$ Wilson coefficient uniquely determines the Veneziano amplitude within the space of physically consistent scattering amplitudes.
Although numerical minimization with some few fixed low-energy data gives results close to Veneziano, our explicit counterexamples show that this proximity does not extend to exact extremality.
Nevertheless, within the standard positivity bootstrap, we have established a conditional theorem showing that the Veneziano amplitude is the unique minimizer when additional UV softness is imposed.

The first counterexample fixes $W_{0,1}=7(\alpha')^4\zeta(4)/4$ and the second additionally fixes $A_{1,0}=(\alpha')^3\zeta(3)$,  yet both allow $A_{0,0}$ to fall below $(\alpha')^2\zeta(2)$ while satisfying the standard bootstrap assumptions and retaining the equally spaced mass-squared spectrum. In particular, their partial-wave positivity holds at every mass level and spin for $4\leq D\leq10$ on a nonzero interval of the deformation parameter.
The failure of exact extremality is therefore not removed by imposing the open-string mass lattice and spin parity.
However, extremality of the leading coefficient does lead to the Veneziano amplitude under the extra assumption of forward ultrasoftness, motivated by the compensating exponential suppression in the string KLT relation. Specifically, with forward softness, we first analytically prove the bound $A_{0,0}\geq (\alpha')^2\zeta(2)$. Then, for $4\leq D\leq10$, we show that equality holds if and only if the amplitude is Veneziano via establishing the string monodromy condition at the inequality saturation.
Neither an equally spaced mass-squared spectrum nor a mass-dependent spin cutoff is imposed on the initial bootstrap class.
We have also linked the minimization of $A_{0,0}$ to the balance of zeros and poles in the full amplitude.

A technical open question is whether fixing more, but still finitely many, Wilson coefficients can establish the uniqueness of the Veneziano amplitude. Based on our construction of explicit counterexamples, we conjecture that finitely many Wilson coefficients combined with leading extremality are insufficient to achieve this uniqueness, since the various parameters in the building blocks of our counterexamples can be adjusted to fine-tune the higher order terms in the amplitude. Furthermore, with a finite number of Wilson coefficients fixed to their Veneziano values, one should be able to establish an upper bound on the deviation from the Veneziano amplitude, as illustrated by the two examples presented in this paper. A broader question is to explore the extremality of $A_{0,0}$ or another Wilson coefficient in a general EFT setup, coupled with weaker UV conditions.

\acknowledgments
We would like to thank Clifford Cheung and Zhuo-Hui Wang for helpful discussions.
SYZ acknowledges support from the National Natural Science Foundation of China under grant No.~12475074 and No.~12247103.

\appendix

\section{Dispersive positivity of counterexamples}
\label{sec:appendix-positivity}

This appendix completes the positivity proof of Section~\ref{sec:positivity} by establishing the uniform upper bound in Eq.~\eqref{eq:difference-bound}.
We use a contour representation of the residues to evaluate the Gegenbauer projection, as in Refs~\cite{Arkani-Hamed:2022gsa,Mansfield:2025gca}.
We follow the general strategy of Ref~\cite{Chen:2026zmz} with suitable modifications, deriving a recursion relation from the Gegenbauer projection and using it to prove the bound in $D=10$.

\subsection{Recursion relation for $h_p$}
\label{sec:appendix-projection}

We derive the recursion relation in Eq.~\eqref{eq:h-recursion} for arbitrary $\nu>0$.
The main steps are to use a contour representation of the residues to reduce the Gegenbauer projection to the Taylor coefficients of a generating function, and then derive the recursion relation for $h_p$ from the differential equation satisfied by this function.

Note that the partial-wave coefficient $g_{N,\ell}^{(\gamma)}$ of the residue $R_N^{(\gamma)}({u})$ is obtained by projection onto the normalized Gegenbauer basis $C_\ell^{(\nu)}(x)/C_\ell^{(\nu)}(1)$, with $x=1+2{u}/N$:
\begin{align}
g_{N,\ell}^{(\gamma)}
&=\frac{C_\ell^{(\nu)}(1)}{\mathfrak h_\ell^{(\nu)}}
\int_{-1}^1(1-x^2)^{\nu-1/2}C_\ell^{(\nu)}(x)
R_N^{(\gamma)}\!\left(\frac N2(x-1)\right)\dd x,\\
&\mathfrak h_\ell^{(\nu)}
\equiv\frac{\pi2^{1-2\nu}\Gamma(\ell+2\nu)}
{\ell!(\ell+\nu)\Gamma(\nu)^2},
\qquad C_\ell^{(\nu)}(1)=\frac{(2\nu)_\ell}{\ell!}.
\end{align}
Using $(1-w)^{-{u}-\gamma-1}=\sum_{k\geq0}({u}+\gamma+1)_k w^k/k!$ and setting $w=1-e^{-2z}=2e^{-z}\sinh z$, we have
\begin{align}
R_N^{(\gamma)}({u}) &=\frac{({u}+\gamma+1)_{N-2\gamma-1}}{(N+\gamma)!} =\frac{(N-2\gamma-1)!}{(N+\gamma)!}
\oint\frac{\dd w}{2\pi\ii}\,
\frac{(1-w)^{-{u}-\gamma-1}}{w^{N-2\gamma}} \nonumber \\
&=\frac{2^{2\gamma-N+1}(N-2\gamma-1)!}{(N+\gamma)!}
\oint\frac{\dd z}{2\pi\ii}\,
\frac{e^{(N+2{u})z}}{(\sinh z)^{N-2\gamma}}.
\end{align}
The contours are small, positively oriented loops around the origin.
Using the Gegenbauer expansion of $e^{a x}$ given in Ref~\cite{Chen:2026zmz}
\begin{align}
e^{ax}=\sum_{\ell\geq0}
\frac{a^\ell}{2^\ell(\nu)_\ell}
{}_0F_1\!\left(;\ell+\nu+1;\frac{a^2}{4}\right)C_\ell^{(\nu)}(x),
\end{align}
and denoting $N=\ell+2L+1$,
we obtain, for $N\geq2\gamma+1$ and $\nu>0$,
\begin{align}
g_{N,\ell}^{(\gamma)}
&=\mathcal N_{N,\ell}^{(\gamma)}h_L^{(\gamma)}(L,\ell),
~~~~
\mathcal N_{N,\ell}^{(\gamma)}
\equiv\frac{2^{2\gamma-N-\ell+1}N^\ell(N-2\gamma-1)!C_\ell^{(\nu)}(1)}{(N+\gamma)!(\nu)_\ell}>0,\\
h_p^{(\gamma)}(L,\ell)&\equiv\oint\frac{\dd z}{2\pi\ii}
\frac{(\sinh z)^{2\gamma}H(z)}{z^{2p+1}},~~~~H(z)\equiv\left(\frac z{\sinh z}\right)^N
{}_0F_1\!\left(;\ell+\nu+1;\frac{N^2z^2}{4}\right) .
\end{align}
To compare $g_{N,\ell}^{(\gamma)}$ with $g_{N,\ell}^{(0)}$, we need to bound $h_L^{(\gamma)}(L,\ell)$ relative to $h_p(L,\ell)\equiv h_p^{(0)}(L,\ell)$.
Since the coefficients $h_p^{(\gamma)}(L,\ell)$ are finite convolutions of $h_p(L,\ell)$ with the known Taylor coefficients of $(\sinh z)^{2\gamma}$, it suffices to control $h_p(L,\ell)$.
We will derive a recursion relation for these coefficients from the differential equation for $H(z)$ and use it to establish the uniform bound in Eq.~\eqref{eq:coefficient-bound}.

Keeping $\nu>0$ arbitrary, we can write the hypergeometric factor as
\begin{align}
\Phi(z)\equiv{}_0F_1\!\left(;\ell+\nu+1;\frac{N^2z^2}{4}\right)=\sum_{k=0}^{\infty}\frac{(N^2z^2/4)^k}{k!(\ell+\nu+1)_k}=\left(\frac{\sinh z}{z}\right)^N H(z).
\end{align}
Its power-series expansion directly yields the differential equation
\begin{align}
z\partial_z^2\Phi+(2\ell+2\nu+1)\partial_z\Phi-N^2z\Phi=0.
\end{align}
Substituting $\Phi(z)=(\sinh z/z)^N H(z)$ into this equation leads to
\begin{align}
&z^2\sinh^2z\,\partial_z^2H
+\left[2Nz^2\sinh z\cosh z+(2\nu-4L-1)z\sinh^2z\right]\partial_zH\nonumber\\
&~~=N\left[(1-N)z^2+(4L+1-2\nu)z\sinh z\cosh z
+(\ell-2L+2\nu-1)\sinh^2z\right]H.
\label{eq:h-ode}
\end{align}
Substituting $H(z)=\sum_{q\geq0}h_q^{(0)}(L,\ell)z^{2q}$ with $h_0^{(0)}=1$ into Eq.~\eqref{eq:h-ode} and matching the coefficient of $z^{2p+2}$ for $p\geq1$,
we get Eq.~\eqref{eq:h-recursion}:
\begin{align}
h_p(L,\ell)&=\sum_{r=2}^{p+1}C_{p,r}(L,\ell)h_{p+1-r}(L,\ell),\qquad p\geq1,
\label{eq:h-recursion-app}\\
C_{p,r}(L,\ell)&=\frac{v_rD_{p,r}(L,\ell)}{4p(p+\ell+\nu)},\\
D_{p,r}(L,\ell)&=N\left[(4L+1-2\nu)r+\ell-2L+2\nu-1\right]\nonumber\\
&\quad-2(p+1-r)\left[2(p+1-r)+2Nr+2\nu-4L-2\right].
\end{align}

\subsection{Uniform bound}
\label{sec:positivity-details}

Here we prove Eq.~\eqref{eq:coefficient-bound} for all $L$ and $\ell$ with $\nu=7/2$ ($D=10$).
For $L=0$, $S_L=0$ and $h_L=1$.
For $L=1,2$, using the recursion relation in Eq.~\eqref{eq:h-recursion-app}, we can easily verify that $S_L<64Nh_L$.
In the following, we shall prove this bound for $L \geq 3$.

For fixed $L$ and $\ell$, $D_{p,r}(L,\ell)$ decreases with $p$ throughout $2\leq r\leq p+1$.
Its signs up to $p=L$ can therefore be determined from the last two rows: $D_{L-1,r}(L,\ell)$ and $D_{L,r}(L,\ell)$.
$D_{L-1,r}(L,\ell)$ is positive for $r\geq2$, and hence
\begin{align}
D_{p,r}(L,\ell)\geq D_{L-1,r}(L,\ell)>0,\qquad1\leq p\leq L-1,\quad2\leq r\leq p+1.
\end{align}
Induction from $h_0=1$ gives $h_p(L,\ell)>0$ for $0\leq p\leq L-1$.
To prove $S_L\leq64N h_L$, however, we need a lower bound on $h_L (L,\ell)$.
In the recursion for $h_L$, the coefficient $C_{L,2}$ can be negative, since its numerator is $D_{L,2}(L,\ell)=\ell(\ell+3)-8(L-1)$.
Since $D_{L,r}(L,\ell)>0$ for $r\geq3$, only the term $C_{L,2}h_{L-1}$ can contribute negatively to $h_L$.

Setting $p=L$ in Eq.~\eqref{eq:h-recursion-app}, we separate the $r=2,3$ terms and write $h_{L-1}=(h_{L-1}-C_{L-1,2}h_{L-2})+C_{L-1,2}h_{L-2}$.
Combining the two contributions proportional to $h_{L-2}$ gives
\begin{align}
    h_{L}=\mathcal C_1(h_{L-1}-\mathcal C_2 h_{L-2})+\mathcal C_3 h_{L-2}+\mathcal R
    \label{eq:hLR001}
\end{align}
where
\begin{align}
\mathcal C_1&=C_{L,2}=\frac{\ell^2+3\ell-8L+8}{6L(2L+2\ell+7)},\qquad
\mathcal C_2=C_{L-1,2}=\frac{\ell^2+11\ell+8L+14}{6(L-1)(2L+2\ell+5)},\label{eq:c12}\\
\mathcal C_3&=C_{L,3}+\mathcal C_1\mathcal C_2=\frac{\ell^2+13\ell+12L+16}{45L(2L+2\ell+7)}+\mathcal C_1\mathcal C_2,\quad
\mathcal R=\sum_{R=3}^L C_{L,R+1}h_{L-R}.
\label{eq:c3-remainder}
\end{align}
Every term in $\mathcal R$ is positive.
To bound $S_L/h_L$, we need to estimate $\mathcal C_1,\mathcal C_2,\mathcal C_3$ and $h_{L-1}-\mathcal C_2h_{L-2}$.
A direct algebraic check of the rational expressions in Eqs.~\eqref{eq:c12}--\eqref{eq:c3-remainder} gives the following bounds for $L\geq3$ and $\ell\geq0$:
\begin{align}
\mathcal C_3>\frac{1}{36L},\qquad \mathcal C_3>\mathcal C_1^2,\qquad \mathcal C_2<40\mathcal C_3.
\label{eq:scalar}
\end{align}
Also, using the definition in Eq.~\eqref{eq:h-recursion-app}, we can get $C_{L-1,R}<\frac32 \sqrt{C_{L,R+1}C_{L-2,R-1}}$.
The recursion relations can be used to express $h_{L-1}-\mathcal C_2h_{L-2}$ and $h_{L-2}$ as sums over the same positive coefficients $h_{L-R}$ appearing in $\mathcal R$.
This gives, upon using $C_{L-1,R}<\frac32 \sqrt{C_{L,R+1}C_{L-2,R-1}}$ and the Cauchy--Schwarz inequality
\begin{align}
h_{L-1}-\mathcal C_2h_{L-2}= \sum_{R=3}^L C_{L-1,R}h_{L-R}\leq\frac32\sqrt{\mathcal R h_{L-2}} .
\end{align}
Using this inequality, Eq.~\eqref{eq:hLR001} and $|\mathcal C_1|<\sqrt{\mathcal C_3}$, we get
\begin{align}
h_L&\geq \mathcal R-\frac32\sqrt{\mathcal C_3 \mathcal R h_{L-2}}+\mathcal C_3 h_{L-2}  =\frac14(\mathcal R+\mathcal C_3 h_{L-2})
+\frac34(\sqrt{\mathcal R}-\sqrt{\mathcal C_3 h_{L-2}})^2
\end{align}
Consequently,
\begin{align}
\frac{\sqrt{\mathcal R h_{L-2}}}{h_L}\leq\frac2{\sqrt{\mathcal C_3}},\qquad\frac{h_{L-2}}{h_L}\leq\frac4{\mathcal C_3},\qquad\frac{\mathcal R}{h_L}\leq4.
\label{eq:RhR80986}
\end{align}

Using the definition in Eq.~\eqref{eq:h-recursion-app}, $D_{L-1,R}(L,\ell)+N(4R-6)\geq2NR(R+1)$ and $4L(L+\ell+7/2)\leq N(N+5)$,
we can derive
\begin{align}
\frac{C_{L,R+1}}{v_R}\geq\frac{12}{7(N+5)}\geq\frac1N\qquad( N\geq7, L \geq R \geq 3).
\end{align}
Since $h_{L-R}>0$, multiplying by $h_{L-R}$ and summing over $R$ yields $\sum_{R=3}^L v_Rh_{L-R}\leq N\sum_{R=3}^L C_{L,R+1}h_{L-R}=N\mathcal R$.
Splitting off the $r=1,2$ terms in Eq.~\eqref{eq:coefficient-bound} then gives
\begin{align}
S_L&=h_{L-1}-\mathcal C_2 h_{L-2}+\Big(\frac13 +\mathcal C_2\Big)h_{L-2}+\sum_{r=3}^Lv_rh_{L-r}\nonumber\\&\leq\frac32\sqrt{\mathcal R h_{L-2}}+\Big(\frac13 +\mathcal C_2\Big)h_{L-2}+N\mathcal R.
\end{align}
Dividing by $h_L$ and using the inequalities in Eqs.~\eqref{eq:RhR80986} and \eqref{eq:scalar}, we obtain
\begin{align}
\frac{S_L}{h_L}\leq4N+\frac3{\sqrt{\mathcal C_3}}+\frac{4(\mathcal C_2+1/3)}{\mathcal C_3}\leq4N+18\sqrt L+48L+160\leq64N.
\end{align}

\section{Elimination of the entire factor}
\label{sec:forward-reconstruction}

Here we prove that the entire factor $e^g$ in $\mathcal G=e^g\mathcal Z/\mathcal P$ is trivial under the assumptions of Section~\ref{sec:minimal-zeros}.
Since $\mathcal G\mathcal P/\mathcal Z$ is entire and has no zeros, it admits an entire logarithm $g$, normalized by $g(0)=0$.
Crossing symmetry then implies that $g$ is even.
As long as we can control the growth of $e^g$, it is not hard to see $g=0$.

To that end, let us define $\mathcal M_F(r)\equiv\max_{|s|\leq r}|F(s)|$. Normalization and the product representation \eqref{eq:zero-and-pole-product} give the following lower and upper bounds for $r>0$:
\begin{align}
0= \log|\mathcal Z(0)| \leq \log \mathcal M_{\mathcal Z}(r) &\leq\max_{|s|\leq r}\frac{1}{2}\sum_{z\in Z(\mathcal G)}\log\left(1+\frac{|s|^2}{|z|^2}\right) = o(r^2),~~~ r\to\infty.
\end{align}
The same argument gives $\log \mathcal M_{\mathcal P}(r)=o(r^2)$. Thus, the summability condition controls the growth of the products. Also, the UV condition controls the growth of the entire function $\mathcal G\mathcal P$.
For each amplitude, we impose forward ultrasoftness at $t=0$ uniformly on a sequence of circles avoiding its poles:
\begin{align}
\Gamma_R=\{s:|s|=R\},\qquad \epsilon_R\equiv\sup_{s\in\Gamma_R}|e^{\ii\pi s}\mathcal A(s,-s)|\longrightarrow0.
\end{align}
Crossing and the maximum modulus principle then give
\begin{align}
\log\mathcal M_{\mathcal G\mathcal P}(R)\leq\log\left(1+\frac{R\epsilon_R}{\pi}\right)+\log\mathcal M_{\mathcal P}(R)=o(R^2).
\end{align}

The behavior of $g$ then follows from $\Re g=\log|\mathcal G\mathcal P|-\log|\mathcal Z|$.
The possible small values of $\mathcal Z$ near its zeros require control of the negative part of its logarithm, which Jensen's formula supplies after angular averaging.
For each $R$, we choose a radius $R/2<\bar{r}<R$ whose circle avoids the common zeros of $\mathcal G\mathcal P$ and $\mathcal Z$; discreteness guarantees its existence, and both logarithmic maximum moduli are $o(\bar{r}^2)$ as $R\to\infty$.
For either $F=\mathcal G\mathcal P$ or $\mathcal Z$, Jensen's formula~\cite[Chapter~5, Theorem~1.1]{stein2003complex} and $F(0)=1$ give a nonnegative angular mean of $\log|F|$:
\begin{align}
\frac{1}{2\pi}\int_0^{2\pi}\log|F(\bar{r}e^{\ii\theta})|\,\dd\theta=\log|F(0)|+\sum_{\substack{a\in Z(F)\\|a|<\bar{r}}}\log\frac{\bar{r}}{|a|}\geq0,
\end{align}
where the sum runs over all zeros of $F$ inside the circle, counted with multiplicity.
The integrated negative part of $\log|F|$ is therefore no larger than its integrated positive part, whose angular mean is bounded by $\log \mathcal M_F(\bar{r})$.
Thus the angular mean of $\bigl|\log|F|\bigr|$ is at most $2\log \mathcal M_F(\bar{r})$, and applying this to both functions gives
\begin{align}
\frac{1}{2\pi}\int_0^{2\pi}|\Re g(\bar{r}e^{\ii\theta})|\,\dd\theta\leq2\log \mathcal M_{\mathcal G\mathcal P}(\bar{r})+2\log \mathcal M_{\mathcal Z}(\bar{r})=o(\bar{r}^2).
\end{align}

Analyticity converts this averaged growth bound into a constraint on every Taylor coefficient of $g$.
Evenness and $g(0)=0$ give $g(s)=\sum_{n\geq1}a_{2n}s^{2n}$.
Multiplying $\Re g(\bar{r}e^{\ii\theta})$ by $e^{-2n\ii\theta}$ and averaging over the circle extracts $a_{2n}\bar{r}^{2n}/2$, so
\begin{align}
\frac{|a_{2n}|\bar{r}^{2n}}{2}\leq\frac{1}{2\pi}\int_0^{2\pi}|\Re g(\bar{r}e^{\ii\theta})|\,\dd\theta=o(\bar{r}^2),\qquad n\geq1.
\end{align}
Dividing by $\bar{r}^{2n}$ and taking $R\to\infty$ gives $a_{2n}=0$ for every $n\geq1$.
In particular, the strict $o(\bar{r}^2)$ estimate eliminates the quadratic term, while evenness and normalization have already excluded the linear and constant terms.
Hence $g=0$ and $\mathcal G=\mathcal Z/\mathcal P$.

\bibliographystyle{JHEP}
\bibliography{draft_jhep}
\end{document}